\RequirePackage[2020-02-02]{latexrelease}
\documentclass[%
reprint,
superscriptaddress,
twocolumn,
showpacs,
amsmath,amssymb,amsthm,amsfonts, 
aps,
pra,
floatfix,notitlepage,latexsym,%longbibliography
]{revtex4-2}
\usepackage{amsmath,amssymb,amsthm,bm,braket,ascmac,bbm,dsfont}
\usepackage{graphicx}
\usepackage{color}
\usepackage{comment}

\newcommand{\aACn}[1]{a_{\mathrm{AC},#1}}
\newcommand{\tACn}[1]{t_{\mathrm{AC},#1}}
\newcommand{\HLZ}{H_\mathrm{LZ}}

\newcommand{\pw}{\nu}
\newcommand{\Hpulse}{H_\mathrm{pulse}}
\newcommand{\Hcw}{H_\mathrm{cw}}
\newcommand{\tin}{t_\mathrm{ini}}
\newcommand{\tfin}{t_\mathrm{fin}}
\newcommand{\Pup}{P_\uparrow}
\newcommand{\mG}{\mathcal{G}}
\newcommand{\mH}{\mathcal{H}}
\newcommand{\mT}{\mathcal{T}}
\newcommand{\mU}{\mathcal{U}}

\newcommand{\adi}{\eta}

\begin{document}
\title{Floquet-Landau-Zener interferometry: Usefulness of the Floquet theory in pulse-laser-driven systems}

%############################################
\author{Tatsuhiko N. Ikeda}
\affiliation{Institute for Solid State Physics, University of Tokyo, Kashiwa, Chiba 277-8581, Japan}
%############################################
\author{Satoshi Tanaka}
\affiliation{Department of Physical Science, Osaka Prefecture University, Sakai, Osaka 599-8531, Japan}
%############################################
\author{Yosuke Kayanuma}
\affiliation{Department of Physical Science, Osaka Prefecture University, Sakai, Osaka 599-8531, Japan}
\affiliation{Laboratory for Materials and Structures, Institute of Innovative Research, Tokyo Institute of Technology, 4259 Nagatsuta, Yokohama 226-8503, Japan}

\date{\today}
\begin{abstract}
We develop the Landau-Zener transfer matrix theory for the instantaneous Floquet states (IFSs) for quantum systems driven by strong pulse laser.
Applying this theory to the pulse excitation probability in two-level quantum systems, we show unexpectedly good quantitative agreements for few-cycle pulses.
This approach enables us to qualitatively understand the probability's peculiar behaviors as quantum path interference between IFSs.
We also study the pulse-width dependence, finding that this Floquet-state interpretation remains useful for shorter pulses down to 2-cycle ones in the present model.
These results imply that the Floquet theory is meaningful in experimental few-cycle lasers if applied appropriately in the sense of IFSs.
\end{abstract}
\maketitle

\section{Introduction}
Intense few-cycle laser fields have opened opportunities for studying strong light-matter interactions and for optically controlling material properties~\cite{Brabec2000}.
Among various approaches, Floquet engineering is an emerging concept in optical material control, in which the time-oscillating nature of fields is utilized~\cite{Bukov2015,Oka2019,Rudner2020}.
The guiding principle is Floquet theory~\cite{Floquet1883,Shirley1965,Sambe1973}, which governs solutions of time-dependent Schr\"{o}dinger equations (TDSE) under perfectly periodic external fields, i.e., infinitely-long pulses.

However, it has not been fully clarified yet when and how Floquet theory is justified under external pulse fields available in experiments.
In this direction, Holthaus and coworkers developed the instantaneous Floquet state (IFS) formalism in their pioneering works~\cite{Breuer1989a,Breuer1989,Drese1999,Holthaus2015}.
Rather than approximating a pulse field crudely by a continuous wave, this formalism utilizes the Floquet states as instantaneous basis states, on which the actual quantum state evolves adiabatically or diabatically during the pulse.
While the theory was elegantly formulated and applied to some examples~\cite{Holthaus2015}, its actual implementation is generically challenging, and its advantages have not been fully explored yet.

In this paper, we further develop the IFS formalism and find the situations where this formalism appropriately describes quantum dynamics under strong pulse fields.
In particular, we introduce the Landau-Zener-type transfer matrices in the Floquet extended Hilbert space, describing multiple transitions between the IFSs quantitatively correctly.
We apply our theory to two-level quantum systems driven by strong pulse fields, showing its applicability and usefulness.
Recent studies showed that the pulse excitation probability of two-level systems exhibits peculiar parameter dependence~\cite{Zhang2017a,Zhang2019,Kayanuma2021b}, but its mechanisms have remained uncovered yet.
Our theory explains even quantitatively that this peculiar behavior is due to quantum path interference between IFSs.
We also study the pulse-width dependence, finding that those Floquet-state interpretations remain valid shorter pulses down to 2-cycle ones in the present model.
These results imply that Floquet theory is meaningful in experimental few-cycle lasers if applied appropriately in the sense of IFSs.

%####
The rest of this paper is organized as follows.
In Sec.~\ref{sec:formulation}, we introduce the pulse excitation problem in a two-level quantum system and demonstrate that the excitation probability exhibits complex behaviors when we vary the pulse strength and the two levels' energy difference. 
To uncover the underlying mechanisms of these behaviors, we review the IFS formalism and develop the Floquet-Landau-Zener (FLZ) theory using transfer matrices for the Floquet Hamiltonian in Sec.~\ref{sec:theory}.
In Secs.~\ref{sec:numerical} and \ref{sec:width}, we implement the FLZ numerically, showing its quantitative success in calculating the pulse excitation probability.
We elucidate that the complex behaviors introduced in Sec.~\ref{sec:formulation} originate from quantum path interference among IFSs.
We also show that these Floquet-based interpretations remain valid for unexpectedly short pulses, including 2-cycle pulse lasers.
%In Sec.~\ref{sec:analytical}, we address a special parameter regime, where we can implement the FLZ theory almost fully analytically and gain deeper insights on the problem.
Finally in Sec.~\ref{sec:summary} we summarize our results and list some open problems for future study.

\section{Formulation of the problem}\label{sec:formulation}

%\subsection{Model and Setup}
For concreteness, we consider an abstract two-level quantum system in strong pulse fields.
Being an effective model in various physical systems, the driven two-level system may represent, e.g., lasing of N${}_2$ molecules~\cite{Zhang2017a,Zhang2019,Kayanuma2021b}, two-band electrons in semiconductors~\cite{Taya2021}, nitrogen-vacancy centers in diamonds~\cite{Fuchs2009}, to name a few.
Throughout this work, we suppose that the Hamiltonian is given by
\begin{align}\label{eq:HtPulse}
    \Hpulse(t) = \frac{b}{2}\sigma_z + a(t) V(t).
\end{align}
Here, $b$ $(>0)$ is the energy difference between the two levels $\ket{\uparrow}$ and $\ket{\downarrow}$.
The coupling to the external field consists of the periodic part $V(t+T)=V(t)$ and the pulse envelope with peak height $a_0$ $(>0)$
\begin{align}
	a(t)=a_0 f(t),	
\end{align}
where $T$ is the period and we define its corresponding angular frequency as $\omega\equiv 2\pi/T$.
We assume that the envelope is positive and normalized so that $\max_t f(t)=1$ and $f(t)\to0$ $(t\to\pm\infty)$.
For concreteness, we focus on the following prototypical coupling term
\begin{align}\label{eq:Vt}
	V(t) = \cos(\omega t) \sigma_x,
\end{align}
which naturally arises in, e.g., two-level atoms coupled to linearly-polarized lasers.
We discuss, in Appendix~\ref{sec:analytical}, modified problems corresponding to circular and elliptic polarizations.
We also specify, for concreteness, the envelope to be Gaussian (generalization to other envelopes is straightforward)
\begin{align}
	%f(t) = e^{-t^2/s^2},
	f(t) = \exp\left[-\left(\frac{t}{\pw T}\right)^2\right],\label{eq:gauss}
\end{align}
where the dimensionless parameter $\pw$ $(>0)$ represents the pulse width in units of $T$.
Namely, we consider a $\nu$-cycle pulse with the envelope~\eqref{eq:gauss}.

%###
Our problem to address is the following.
Suppose that our initial state at $t=-\infty$ is the ground state
$\ket{\Psi(-\infty)}=\ket{\downarrow}$,
which evolves in time according to the time-dependent Schr\"{o}dinger equation (TDSE)
\begin{align}\label{eq:Schpulse}
i\frac{d}{dt}\ket{\Psi(t)}=\Hpulse(t)\ket{\Psi(t)}	
\end{align}
to become
\begin{align}
	\ket{\Psi(+\infty)} = \exp_+\left(-i\int_{-\infty}^\infty \Hpulse(\tau)d\tau\right) \ket{\downarrow}
\end{align}
after the pulse
($\hbar=1$ throughout this paper).
Then, we ask the final weight of the excited state
\begin{align}
	P_\uparrow = | \braket{\uparrow|\Psi(+\infty)}|^2.
\end{align}

%#####################
\begin{figure}
	\includegraphics[width=\columnwidth]{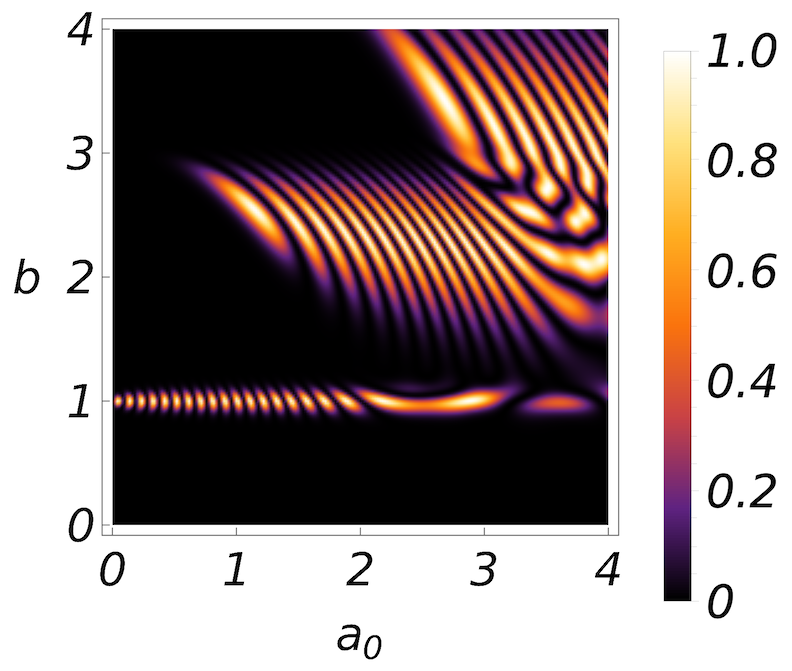}
	\caption{Excitation probability $P_\uparrow$ plotted against the pulse peak height $a_0$ and the energy level difference $b$ with pulse width $\pw=6$.}
	\label{fig:mask}
\end{figure}

%####
Despite this simple setup, $P_\uparrow$ variously changes depending on the energy level difference $b$, coupling energy $a$, driving frequency $\omega$, and pulse width $s$.
We show in Fig.~\ref{fig:mask} the results of $\Pup$ in terms of the numerical integration of TDSE~\eqref{eq:Schpulse} with frequency $\omega=1$ and pulse width $\pw=6$.
At weak coupling $a_0$, $P_\uparrow$ becomes significant only near resonance $b/\omega\simeq1$.
This parameter regime is well described by the rotating-wave approximation, and the oscillating behavior of $\Pup$ is understood in relation to the Rabi oscillation~\cite{Rabi1937} (see also Appendix~\ref{sec:circ}).

%####
Furthermore, away from the resonance and at strong couplings, there is a complex pattern of the region for nonvanishing $\Pup$.
For example, as $a_0$ increases with $b=2.5$ held fixed, $\Pup$ suddenly grows up at $a_0\sim 1.0$, shows clear oscillations up to $a_0\sim3.0$, and then exhibits irregular behaviors $a_0\gtrsim3.0$.
The sudden grow-up was shown, in the pioneering works by Holthaus and coworkers~\cite{Holthaus2015}, to be due to avoided crossing structures of Floquet quasienergies as we will review below in Sec.~\ref{sec:IFS}.
However, the oscillations and irregular behaviors have not been studied well.
In the following, we extend their theory by combining the Landau-Zener transfer matrix and elucidating those complex behaviors of $\Pup$ in the whole parameter region.

%#######################################################
\section{Landau-Zener-St\"{u}ckelberg theory for Floquet states}\label{sec:theory}

\subsection{Instantaneous Floquet states (IFS)}\label{sec:IFS}
The key to understanding the complex dynamics is using the basis of the instantaneous Floquet states (ITS)~\cite{Breuer1989,Holthaus2015},
which we briefly review here.
We note that this formulation was generalized, in Ref.~\cite{Drese1999}, to the case where $\omega$ also varies slowly.

The Floquet states are defined by the solutions to the time-dependent Schr\"{o}dinger equation for the continuous wave rather than the pulse.
Namely, according to Floquet theory~\cite{Shirley1965,Sambe1973}, the two independent solutions to
\begin{align}\label{eq:Schcw}
i\frac{d}{dt}\ket{\psi(t)}=\Hcw(a,t)\ket{\psi(t)}	
\end{align}
with
\begin{align}\label{eq:Htcw}
	\Hcw(a,t)= \frac{b}{2}\sigma_z + a V(t)
\end{align}
can be written in the following forms:
\begin{align}\label{eq:Floquetdef}
	\ket{\psi_m(a,t)} = e^{-i \epsilon_m(a)t}\ket{u_m(a,t)}\quad (m=1,2).
\end{align}
Note that $\Hcw(a,t)$ is obtained by replacing the envelope $a(t)$ in $\Hpulse(t)$ by a constant $a$ (i.e., replacing the Gaussian pulse by a continuous wave).
In Eq.~\eqref{eq:Floquetdef}, $\ket{u_m(a,t)}=\ket{u_m(a,t+T)}$ are periodic and called the Floquet states, and the real numbers $\epsilon_m(a)$ are quasienergies.
We explicitly put the dependence on the coupling strength $a$ on the Floquet states and quasienergies that will play crucial roles.

\begin{figure}
	\includegraphics[width=\columnwidth]{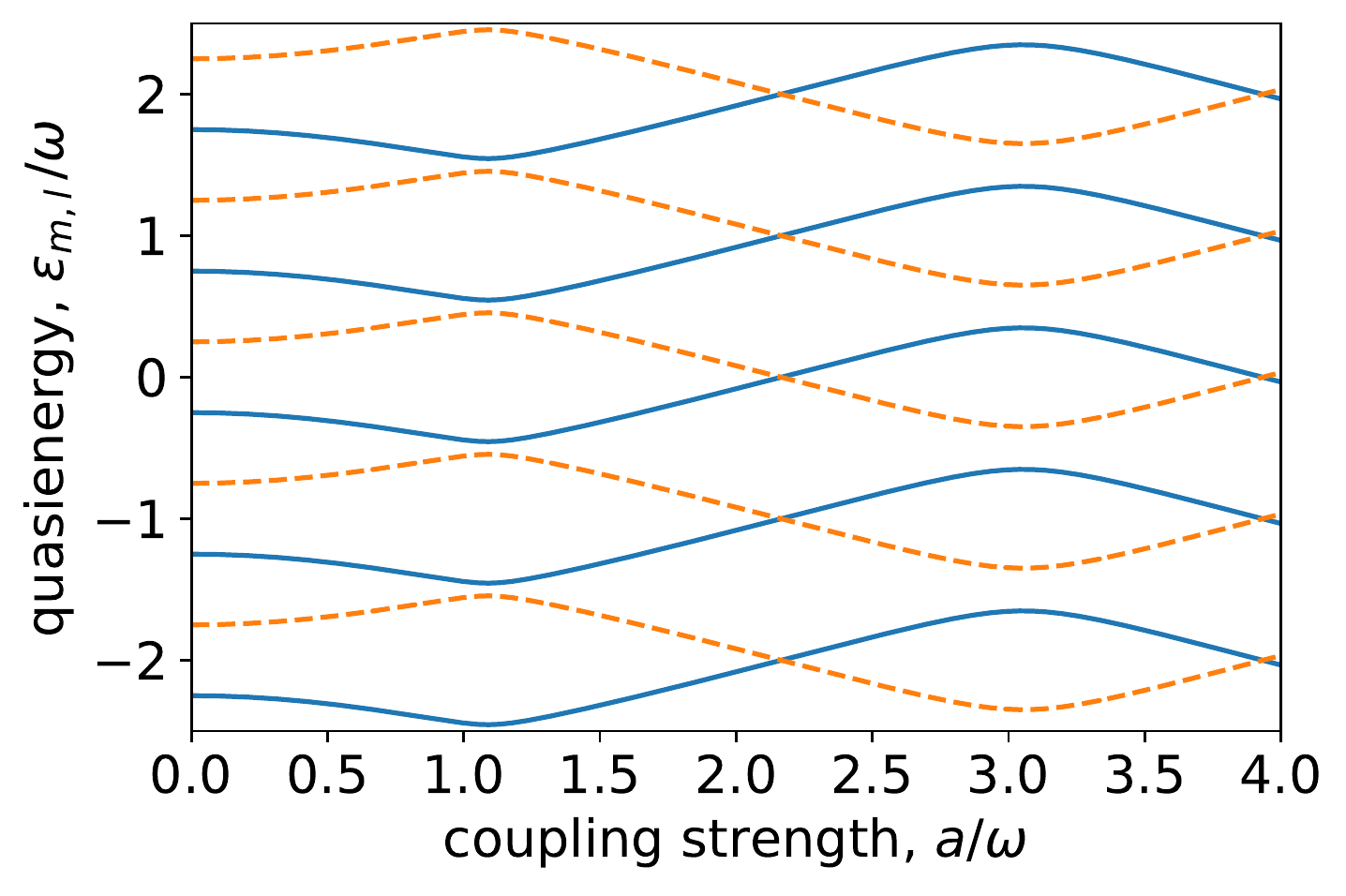}
	\caption{Quasienergies for $b=2.5$ plotted against coupling strength $a$. Solid (dashed) lines show those
	for Floquet states of $m=2$ ($m=1$) approaching $\ket{\downarrow}$ ($\ket{\uparrow}$) as $a\to0$.
	Five Floquet replicas are presented.}
	\label{fig:quasi}
\end{figure}

%###
We remark the famous replicas of Floquet states.
Note that Eq.~\eqref{eq:Floquetdef} can also be written as
$\ket{\psi_m(a,t)} = e^{-i [\epsilon_m(a)+l\omega]t}e^{il\omega t}\ket{u_m(a,t)}$
for an arbitrary integer $l\in\mathbb{Z}$.
Being periodic,
\begin{align}\label{eq:defu}
	\ket{u_{m,l}(a,t)} \equiv e^{il\omega t}\ket{u_m(a,t)}\quad(m=1,2)
\end{align}
are all Floquet states, and their quasienergies are given by
\begin{align}\label{eq:quasi_replica}
	\epsilon_{m,l}(a) \equiv \epsilon_m(a) +l\omega.
\end{align}
In Fig.~\ref{fig:quasi}, we plot the quasienergies with replicas numerically obtained for $b=2.5$.
They show avoided crossings near $a=1.0$ and $3.0$, where two quasienergies repel each other.
This is a manifestation of strong hybridization between $\ket{\uparrow}$ and $\ket{\downarrow}$, and we will discuss, in detail, how this hybridization leads to the complex pattern in Fig.~\eqref{fig:mask}.
We remark that the quasienergies of $m=1$ and $2$ do not repel but cross near $a=2.2$, which is due to a selection rule prohibiting hybridization (see Appendix~\ref{sec:elliptic}).

Although these replicas lead to the same solution to the Schr\"{o}dinger equaion~\eqref{eq:Schcw},
they are all necessary when one expands an arbitrary periodic function $F(t)$ with Floquet states.
In other words, the replicas satisfy the completeness relation
\begin{align}\label{eq:identity}
	\sum_{m=1,2}\sum_{l=-\infty}^\infty \ket{u_{m,l}(a,t)}\bra{u_{m,l}(a,t')} = T\delta_T(t-t')I,
\end{align}
where $\delta_T(t) \equiv \sum_{n=-\infty}^\infty \delta(t-nT)$ and $I$ is the identity operator.

%#####
The IFS formalism is to expand the solution for the pulse problem $\ket{\Psi(t)}$  in terms of the Floquet states:
\begin{align}\label{eq:IFSexpansion}
	\ket{\Psi(t)} = \sum_\alpha c_\alpha(t) \ket{u_\alpha(a(t),t)},
\end{align}
where $\alpha=(m,l)$ is a shorthand notation for the double indices.
Substituting Eq.~\eqref{eq:IFSexpansion} into Eq.~\eqref{eq:Schpulse}, we have the time-evolution equation for the expansion coefficients (see Appendix~\ref{sec:c_eq}),
\begin{align}\label{eq:IFSevol}
	i\frac{dc_\alpha(t)}{dt} &= \sum_\beta \mathcal{H}_{\alpha\beta}(a(t)) c_\beta(t),
\end{align}
where $\mathcal{H}_{\alpha\beta}(a(t))$ is the infinite-dimensional ``Hamiltonian'' defined by
\begin{align}
	\mathcal{H}_{\alpha\beta}(a(t)) &\equiv \delta_{\alpha\beta}\epsilon_\alpha(a(t))-i\frac{da}{dt}\mathcal{G}_{\alpha\beta}(a(t)),\label{eq:FHam}\\
	\mathcal{G}_{\alpha\beta}(a) &\equiv\int_0^T \frac{dt}{T}\braket{u_\alpha(a,t) |\partial_a| u_\beta(a,t)}.\label{eq:defG}
\end{align}
Here, $\partial_a\equiv \partial/\partial a$, and we have assumed that $\ket{u_\alpha(a,t)}$ are differentiable for $a$ by requiring the gauge-fixing condition $\braket{u_\alpha(a,t)|\partial_a |u_\alpha(a,t)}=0$.
Equation~\eqref{eq:FHam} means that the ``Hamiltonian'' in the extended (Sambe~\cite{Sambe1973}) space has the quasienegies in its diagonal elements, and $\mG_{\alpha\beta}$ causes transitions between the Floquet states.

We remark an ambiguity in expanding the physical state $\ket{\Psi(t)}$ in terms of the Floquet replica index $l$.
To work in the IFS, we fix the initial values of $c_\alpha(t)$ by $\ket{\Psi(\tin)}=\sum_\alpha c_\alpha(\tin)\ket{u(a(\tin),\tin)}$, which has an infinite number of solutions due to the Floquet replicas.
However, when we calculate physical observables such as $|\braket{\uparrow|\Psi(\tfin)}|^2$,
the results do not depend on which initial condition is used~\cite{Drese1999}.
Intuitively, this independence is based on the fact that the ambiguity happens only between the physically-equivalent states.
Thus, in the following, we assume that only one $l$ is weighted in the initial condition.

%####
This formalism helps us to interpret physical results.
Following Ref.~\cite{Holthaus2015},
let us interpret how $\Pup$ suddenly grows up at $a_0\sim1.0$ as $a_0$ increases from zero at, e.g., $b=2.5$.
Initially $(t\to -\infty)$, our state is $\ket{\downarrow}$, and the Floquet states there coincide with the energy eigenstates $\ket{\uparrow}$ and $\ket{\downarrow}$ since the coupling vanishes $a(t=-\infty)=0$.
Thus, we can set the initial state in the extended space as $c_{m=2,l=0}(-\infty)=1$ and $c_\alpha(-\infty)=0$ for $\alpha\neq(2,0)$.
Then, this state evolves according to Eq.~\eqref{eq:IFSevol}, where the coupling envelope $a(t)$ slowly varies.
Graphically, our initial state lies at a single left-end of a solid line in Fig.~\ref{fig:quasi}, and it goes right as $a$ increases with time.
This evolution is adiabatic and transitions between $\alpha$'s are unlikely as long as $-i\frac{da}{dt}\mG_{\alpha\beta}(a(t))$ in Eq.~\eqref{eq:FHam} is negligible.
For slowly varying $a(t)$, this condition breaks down at the first avoided crossing point $a=\aACn{1}\sim1.0$, where a part of the wave function is transferred to another state represented by a dashed line.
Therefore, for $a_0<\aACn{1}$, the whole dynamics is adiabatic, the final state is almost the same as the initial state, and $\Pup\sim0$.
Meanwhile, for $a_0>\aACn{1}$, the state experiences transitions twice, the final state is a superposition of the states on the solid and dashed lines in Fig.~\ref{fig:quasi}, and $\Pup$ becomes nonvanishing.
This is the IFS interpretation for the sudden increase of $\Pup$ along e.g. $b=2.5$ in Fig.~\ref{fig:mask}.

%##########################################
\subsection{Transfer Matrices}\label{sec:transfer}
One is naturally led to the question: Does the IFS viewpoint allow us to understand the whole complex structure in Fig.~\ref{fig:mask}?
To the authors' knowledge, although the Landau-Zener-like transition probability at a single passage of an avoided crossing was analyzed~\cite{Drese1999}, the interference pattern has not been well studied.
Our aim is to introduce the Landau-Zener-St\"{u}ckelberg transfer matrix method in the extended space and to show that the IFS formalism is very powerful even quantitatively.

As we discussed at the end of Sec.~\ref{sec:IFS}, the ``Hamiltonian'' $\mH(a(t))$ depends on time through the envelope $a(t)$, and its eigenvalues (quasienergies) form avoided crossings.
In applying the transfer matrix method to the IFS (see Appendix~\ref{sec:LZ} for this method in the conventional sense), we need two generalizations: (i) there are quasienergy replicas of avoided crossings and (ii) the system passes avoided crossings multiple times in $-\infty<t<\infty$ when the pulse peak $a_0$ is large.

%####
Suppose that there are $N$ $(>0)$ avoided crossing points denoted by $\{\aACn{n}\}_{n=1}^N$ below the pulse peak height $a_0$ and they are in ascending order: $0<\aACn{1}<\aACn{2}<\dots<\aACn{N}<a_0$.
For example, we have $N=2$ for $a_0=3.5$ in Fig.~\ref{fig:quasi}.
Correspondingly, we define the crossing times $\tACn{n}$ $(>0)$ by
\begin{align}
	a(\tACn{n})=a_0f(\tACn{n})=\aACn{n}\quad (n=1,2,\dots,N).
\end{align}
For simplicity, we assume that the envelope is even, $f(-t)=f(t)$,
and monotonically decreasing in $t\ge0$ as the Gaussian envelope is.
Then, the crossings happen also at $t=-\tACn{n}$ $(n=1,2,\dots,N)$, and we have
\begin{align}
	-\tACn{1}<\dots<-\tACn{N}<0<\tACn{N}<\dots <\tACn{1}.
\end{align}

At the $n$-th crossing point, the transfer matrix $\mT_n$, which will be defined in Eq.~\eqref{eq:Tmat}, connects the state vectors before and after the crossing as
\begin{align}
	\vec{c}\,(t=\tACn{n}^+) = \mT_n \vec{c}\,(t=\tACn{n}^-).
\end{align}
Here $\vec{c}(t)$ is the vector notation for $c_\alpha(t)$'s, $\tACn{n}^\pm\equiv \tACn{n}\pm0$ (we will also use $-\tACn{n}^\pm\equiv -\tACn{n}\pm0$), and $\mT_n$ is an infinite-dimensional matrix given as follows.
The avoided crossing occurs between a pair of Floquet states, which we label as $(m_U,l_U)$ and $(m_L,l_L)$.
Here, the subscript $U$ ($L$) denotes the upper (lower) levels at the crossing.
For example, $(m_U,m_L)=(2,1)$ and $(1,2)$ at the first and second crossings, respectively, in Fig.~\ref{fig:quasi}.
In this notation, the nonzero matrix elements of $\mT_n$ are given as
\begin{align}
	(\mT_n)_{\alpha\beta} &= 
	\begin{pmatrix}
		\sqrt{1-P_n}e^{-i\varphi^S_n} & -\sqrt{P_n} \\
		\sqrt{P_n} & \sqrt{1-P_n} e^{i\varphi^S_n}
	\end{pmatrix}_{\alpha\beta},\label{eq:Tmat}
\end{align}
where $\alpha=(m_U,l_U)$ and $(m_L,l_L)$ correspond to the first and second rows, respectively.
The two parameters $P_n$ and $\varphi_n^S$ are the Landau-Zener transition probability and the Stokes phase for the $n$-th avoided crossing (see Sec.~\ref{sec:LZ}),
\begin{align}
	P_n &= \exp(-2\pi \delta_n),\label{eq:Pn}\\
	\delta_n &= \frac{\Delta_n^2}{4v_n},\label{eq:deltan}\\
	\varphi_n^S &= -\frac{\pi}{4}+\delta_n \ln(\delta_n-1) + \arg\Gamma(1-i\delta_n).\label{eq:phisn}
\end{align}
Here, $\Delta_n$ and $v_n$ are the quasienergy gap and the passing speed at the $n$-th avoided crossing, respectively.
These parameters are defined in the approximate form of the pair quasienergies near $t=\tACn{n}$
\begin{align}
	\epsilon_\alpha(a(t)) &\simeq \mathrm{const.} \pm \sqrt{ \left(\frac{\Delta_n}{2}\right)^2 +\left[\frac{v_n(t-\tACn{n})}{2} \right]^2 }\\
	&\simeq \mathrm{const.} \pm \left[\frac{\Delta_n}{2}+ \frac{v_n^2(t-\tACn{n})^2}{4\Delta_n}\right]\label{eq:quasi_expansion}
\end{align}
for $\alpha=(m_U,l_U)$ and $(m_L,l_L)$.
Note that $\Delta_n$ and $v_n$ are well-defined in that they are the same for every Floquet replica.

To obtain $\Delta_n$ and $v_n$ in practice, we expand $\epsilon_\alpha(a)$ around $a=\aACn{n}$.
Since $d\epsilon_\alpha(a)/da$ vanishes at the avoided crossing, we have the following second-order series expansion 
$\epsilon_\alpha(a(t))\simeq \epsilon_\alpha(\aACn{n})+\frac{1}{2}(d^2 \epsilon_\alpha(a)/da^2)(da/dt)^2(t-\tACn{n})^2$,
where $d^2\epsilon_\alpha(a)/da^2$ and $da/dt$ are evaluated at $a=\aACn{n}$ and $t=\tACn{n}$, respectively.
Comparing this with Eq.~\eqref{eq:quasi_expansion}, we have
\begin{align}
\Delta_n &= \epsilon_{(m_U,l_U)}(\aACn{n})-\epsilon_{(m_L,l_L)}(\aACn{n}),\\
%v_n&=\sqrt{2\Delta_n|(d^2 \epsilon_\alpha(a)/da^2)|}|da/dt|,
v_n&=\sqrt{2\Delta_n\left|\frac{d^2 \epsilon_\alpha(a)}{da^2}\right|}\left|\frac{da}{dt}\right|,
\end{align}
where $\alpha$ is either $(m_U,l_U)$ or $(m_L,l_L)$ that give the same $|d^2 \epsilon_\alpha(a)/da^2|$.
One can obtain these parameters by numerical fitting as we will implement in Sec.~\ref{sec:numerical} or by analytical calculations for some special cases as we will demonstrate in Appendix~\ref{sec:analytical}.

Except for the crossing points, the evolution is assumed to be mere phase acquisitions due to the first term on the right-hand side of Eq.~\eqref{eq:FHam}.
In the vector notation, we have
\begin{align}
	\vec{c}\,(t=\tACn{n}^-) = \mU_{n,n+1}\vec{c}\,(t=\tACn{n+1}^+),
\end{align}
where $\mU_{n,n+1}$ is diagonal and
\begin{align}\label{eq:Umat}
	(\mU_{n,n+1})_{\alpha\alpha} = \exp\left[ -i\int_{\tACn{n+1}}^{\tACn{n}}ds\epsilon_\alpha(a(s))\right].
\end{align}
For convenience, we define $(\mU_{N,N+1})_{\alpha\alpha} = \exp\left[ -i\int_0^{\tACn{N}}ds\epsilon_\alpha(a(s))\right]$.

Since we are considering a symmetric envelope $a(-t)=a(t)$, time evolution is symmetric in $-\infty< t \le 0$ and $0\le t< \infty$.
The transfer matrix $\mT_n$ describes the state transfer both at $t=\pm \tACn{n}$, and
$\mU_{n+1,n}$ represents the phase acquisition not only from $\tACn{n+1}$ to $\tACn{n}$ but also from $-\tACn{n}$ to $-\tACn{n+1}$.
Thus, we obtain the state transfer between the first and final avoided crossings,
\begin{widetext}
\begin{align}\label{eq:cevol}
	\vec{c}\,(t=\tACn{1}^+)=  \left[\prod_{n=N}^{1} \mT_n \mU_{n,n+1}\right]\left[\prod_{n=1}^{N} \mU_{n,n+1}\mT_n\right] \vec{c}\,(t=-\tACn{1}^-).
\end{align}
\end{widetext}
The entire dynamics is obtained by the phase acquisitions before (after) the first (final) avoided crossing:
$\vec{c}(t=-\tACn{1}^-)= \mU_{<} \vec{c}(t=\tin)$ and $\vec{c}(t=\tfin)= \mU_{>} \vec{c}(t=\tACn{1}^+)$, where $(\mU_{<})_{\alpha\alpha} = \exp\left[ -i\int_{\tin}^{-\tACn{1}}ds\epsilon_\alpha(a(s))\right]$ and $(\mU_{>})_{\alpha\alpha} = \exp\left[ -i\int_{\tACn{1}}^{\tfin}ds\epsilon_\alpha(a(s))\right]$ with $\tin=-\infty$ and $\tfin=+\infty$.

%###
Thus, we have obtained the whole evolution of wave vector $\vec{c}(t)$ in the IFS based on the transfer matrix method.
The physical interpretation is clear in Eq.~\eqref{eq:cevol}.
The wave vector experiences adiabatic dynamics described by the phase factors $\mU_{n,n+1}$ and Landau-Zener-like diabatic dynamics described by the transfer matrices $\mT_n$.
The phase factors due to the Stokes phase in $\mT_n$ and $\mU_{n,n+1}$ amount to the St\"{u}ckelberg phase and cause interferences as we will see in the following.

%###
Finally, we formulate how to calculate the physical observable $\Pup$ of interest from $\vec{c}$.
By using Eq.~\eqref{eq:IFSexpansion}, we have
\begin{align}
\Pup &=|\braket{\uparrow|\Psi(\tfin)}|^2\\
&=\left|\sum_\alpha c_\alpha(\tfin)\braket{\uparrow|u_\alpha(a(\tfin),\tfin}\right|^2.	\label{eq:PupC1}
\end{align}
We recall Eq.~\eqref{eq:defu} and suppose that $\tfin\to \infty$, in which $\ket{u_m(a(\tfin)),\tfin}\to \ket{u_m(0,\infty)}=\delta_{m1}\ket{\uparrow}+\delta_{m2}\ket{\downarrow}$.
Thus, the sum over $\alpha=(m,l)$ in Eq.~\eqref{eq:PupC1} is trivially taken for $m$, and we have
\begin{align}
	\Pup = \left|\sum_l c_{1,l}(\tfin)e^{il\omega \tfin}\right|^2.\label{eq:PupC2}
\end{align}
Note that $\vec{c}(\tfin)$ is connected to $\vec{c}(\tACn{1}^+)$ by
$c_{m,l}(\tfin)
=\exp[-i\int_{\tACn{1}}^{\tfin} \epsilon_{m,l}(a(s))ds] c_{m,l}(\tACn{1}^+)
=\exp[-i\int_{\tACn{1}}^{\tfin} \epsilon_{m}(a(s))ds]e^{-il\omega (\tfin-\tACn{1})} c_{m,l}(\tACn{1}^+)$, where we have used Eq.~\eqref{eq:quasi_replica}.
Substituting this equation into Eq.~\eqref{eq:PupC2}, we obtain
\begin{align}
	\Pup = \left|\sum_l c_{1,l}(\tACn{1}^+)e^{il\omega \tACn{1}}\right|^2.\label{eq:PupC3}
\end{align}
Equation~\eqref{eq:PupC3} is useful since we can compute the excitation probability $\Pup$ just after the final passage of the avoided crossing, and $\vec{c}(\tACn{1}^+)$ is given in Eq.~\eqref{eq:cevol}.

We remark that, in Eq.~\eqref{eq:cevol}, 
we can set $c_{m=2,l=0}(-\tACn{1}^-)=1$ and $c_\alpha(-\tACn{1}^-)=0$ for $\alpha\neq(2,0)$.
These conditions are what we imposed for $t=\tin$ at the end of Sec.~\ref{sec:IFS}.
Nonetheless, the evolution between $t=\tin$ and $-\tACn{1}$ merely gives an overall phase factor, which is irrelevant for $\Pup$.

%%%
To summarize the transfer matrix method for the IFS, our recipe for obtaining $\Pup$ consists of using $\vec{c}(-\tACn{1}^-)$ thus specified, transferring the state by Eq.~\eqref{eq:cevol}, and invoking Eq.~\eqref{eq:PupC3}.
While this method is an approximation, its physical interpretation is clear in that the evolution is a close analog of the Landau-Zener-St\"{u}ckelberg interferometry on the Floquet states.
In the following, we will implement this recipe and show that it works well even quantitatively.

%%%%%
\section{Numerical implementation}\label{sec:numerical}
In this section, we apply the transfer matrix method to understand the complex structure in Fig.~\ref{fig:mask}.
We will focus on 6-cycle pulses ($\nu=6$), which are so long that the transfer matrix method works well. We will discuss how results change with the pulse width $\nu$ later in Sec.~\ref{sec:width}.

\begin{figure}
	\includegraphics[width=\columnwidth]{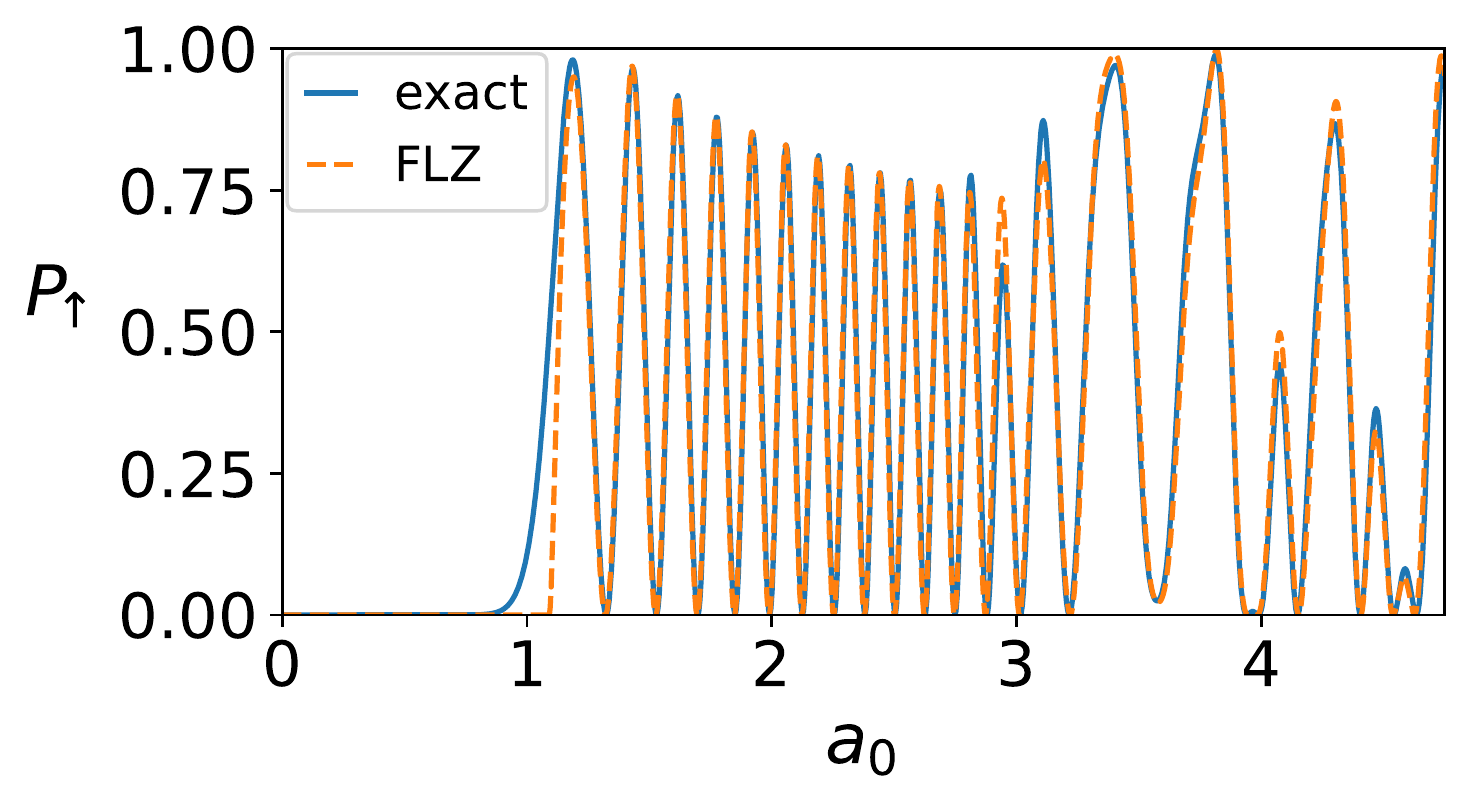}
	\caption{Excitation probability $P_\uparrow$ for $b=2.5$ plotted against the pulse peak height $a_0$ for the pulse width $\pw=6$. The solid and dashed curves show $P_\uparrow$ obtained, respectively, by solving the TDSE~\eqref{eq:Schpulse} numerically and by invoking the FLZ theory~\eqref{eq:PupC3}.}
	\label{fig:PLZ}
\end{figure}

As shown in Sec.~\ref{sec:transfer}, the necessary information to implement the method are all obtained from the quasienergies plotted in Fig.~\ref{fig:quasi}.
To be specific, we set $b=2.5$, for which $\aACn{1}=1.09$ ($\Delta_1=8.98\times10^{-2}$) and $\aACn{2}=3.05$ ($\Delta_2=0.302$) are obtained numerically, from which we can calculate the $a_0$-dependence of $\Pup$ by the Floquet-Landau-Zener (FLZ) theory~\eqref{eq:PupC3}.
In Fig.~\ref{fig:PLZ}, we compare $\Pup$ for $b=2.5$ and $\nu=6$ obtained by the exact numerical simulation of the TDSE~\eqref{eq:Schpulse} and by the FLZ method, where the results are shown for $0\le a_0 \le \aACn{3}=4.75$.

%###
For $a_0<\aACn{1}$, the transfer matrix approach tells us that there is no state transfer between the Floquet states and hence $\Pup=0$ as shown in Fig.~\ref{fig:PLZ}.
This result agrees with $\Pup$ obtained directly by the TDSE for $a_0$ well below $\aACn{1}$.
Near $a_0=\aACn{1}$, the transfer matrix deviates from the exact result.
This deviation originates from the adiabatic-impulse approximation in that the state transfer occurs instantaneously right at the avoided crossing and is a close analog of the deviation in the conventional Landau-Zener problem explained in Appendix~\ref{sec:LZ}.
Except for $a_0\simeq \aACn{n}$ $(n=1,2,\dots)$, we expect that the transfer matrix method works well.

\begin{figure}
	\includegraphics[width=\columnwidth]{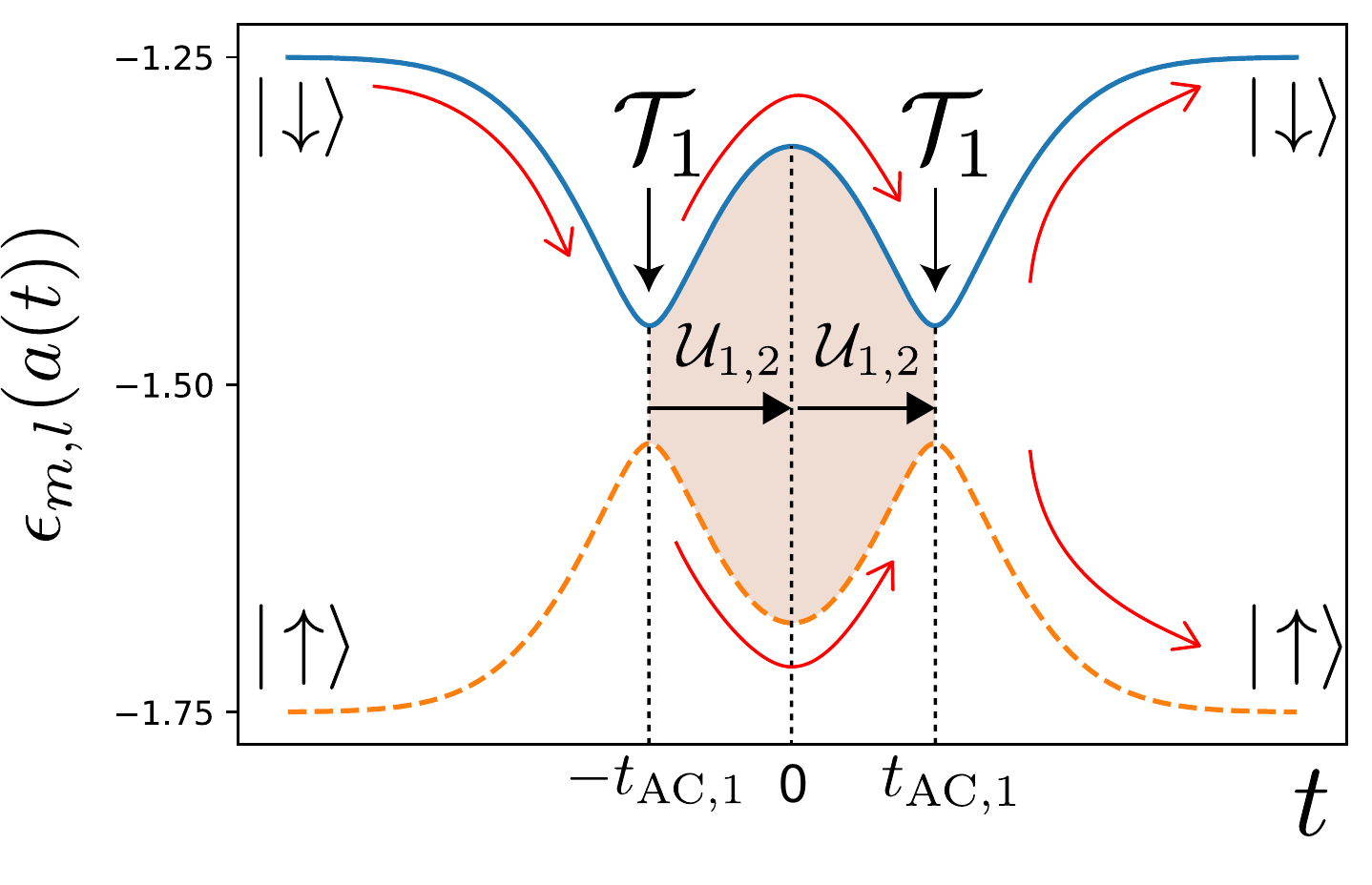}
	\caption{Schematic illustration of the Floquet-Landau-Zener interferometry for $\aACn{1}<a_0<\aACn{2}$.
	The curved arrows show quantum state trajectories along the IFS, and the two $\mU_{1,2}$'s produce a relative phase factor between the upper and lower trajectories between the two LZ transitions denoted by the transfer matrices $\mT_1$.}
	\label{fig:quasi_diagram}
\end{figure}

%####
For $\aACn{1}<a_0<\aACn{2}$, we see the essence of the Landau-Zener-St\"{u}ckelberg interferometry for the IFS.
For this case, there are two passages of the same avoided crossing point $\aACn{1}$, and we have
\begin{align}\label{eq:cevol1}
	\vec{c}\,(t=\tACn{1}^+)=   \mT_1 (\mU_{1,2})^2 \mT_1\vec{c}\,(t=-\tACn{1}^-),
\end{align}
which follows from Eq.~\eqref{eq:cevol}.
The physical interpretation of Eq.~\eqref{eq:cevol1} is schematically illustrated in Fig.~\ref{fig:quasi_diagram}.
At the first crossing $t=-\tACn{1}$, a superposition of two Floquet states $(m,l)=(2,0)$ and $(1,-3)$ is created by $\mT_1$, and these state acquire phase factors due to $(\mU_{1,2})^2$ until $t=\tACn{1}$.
At the second crossing $t=\tACn{1}$, the superposed states experience the state mixing again by $\mT_1$, and the final state has nonvanishing weight on $(m,l)=(1,-3)$, which adiabatically approaches $\ket{\uparrow}$ as $t\to +\infty$.

Since there are only two Floquet states are involved in this case, we can simplify Eq.~\eqref{eq:cevol1} by ignoring irrelevant zero elements.
Focusing on the subspace for $(m,l)=(2,0)$ and $(1,-3)$, we set $\vec{c}={}^t(1,0)$, for which Eqs.~\eqref{eq:Tmat} and \eqref{eq:Umat} give $c_{1,-3}(t=\tACn{1}^+)=\sqrt{P_1(1-P_1)}[e^{i(\varphi_1^S-2\Phi_{1,-3})}+e^{i(-\varphi_1^S-2\Phi_{2,0})}]$
with
\begin{align}
\Phi_\alpha \equiv \int_{0}^{\tACn{1}}ds\,\epsilon_\alpha(a(s)).\label{eq:Phi}	
\end{align}
Thus, from Eq.~\eqref{eq:PupC3}, we obtain
\begin{align}\label{eq:PupN1}
	\Pup = 4P_1(1-P_1)\cos^2(\varphi_1^S+\Phi_{2,0}-\Phi_{1,-3}).
\end{align}
Here, the quantum path interference effect in $\Pup$ is evident, and the phase $\varphi_1^S+\Phi_{2,0}-\Phi_{1,-3}$ is a Floquet generalization of the St\"{u}ckelberg phase~\cite{Shevchenko2010}.

%###
The $a_0$ dependence of $\Pup$ calculated from Eq.~\eqref{eq:PupN1} well describes that obtained numerically exactly for $\aACn{1}<a_0<\aACn{2}=3.05$ as shown in Fig.~\ref{fig:PLZ}.
Let us discuss two characteristic behaviors of $\Pup$ in this region: (i) $\Pup$ oscillates, and 
(ii) the envelope of $\Pup$ quickly increases in $\aACn{1}<a_0\lesssim 1.2$ and then slowly decreases in $1.2\lesssim a_0 <\aACn{2}$.
The first character (i) is mainly due to $\Phi_{2,0}-\Phi_{1,-3}$, which is the integrated phase difference between the Floquet states and corresponds to half the area of the shaded region in Fig.~\ref{fig:quasi_diagram}.
As $a_0$ increases, $\tACn{1}$ increases, and so does $\Phi_{2,0}-\Phi_{1,-3}$.
Inside the cosine (see Eq.~\eqref{eq:PupN1}), the increase of $\Phi_{2,0}-\Phi_{1,-3}$ results in the oscillating behavior of $\Pup$.
The second character (ii) is due to $P_1(1-P_1)$ in Eq.~\eqref{eq:PupN1}.
We recall that $P_1$ depends on $a_0$ only through $da/dt|_{t=\tACn{1}}$ in the crossing speed $v_1$.
Since our envelope is Gaussian, as $a_0$ increases from $\aACn{1}$, $\tACn{1}$ does from zero.
During this, $da/dt|_{t=\tACn{1}}$ first increases and then decreases.
This nonmonotonic behavior results in the character (ii) through $\delta_1$ and hence $P_1$.

%####
While we have focused on $b=2.5$, the interpretation by two passages of avoided crossings also apply to other points in Fig.~\ref{fig:mask}.
For example, roughly in the region $(a_0,b)\in [1,3]\times[1.5,3]$ as well as $(a_0,b)\in [3,4]\times[3,4]$, we see regular patterns of curves, in which $\Pup$'s behavior follows from similar mechanisms illustrated in Fig.~\ref{fig:quasi_diagram}.

%###
Now, we come back to $b=2.5$ and consider $\aACn{2}<a_0<\aACn{3}=4.75$ to elucidate the complex pattern in $(a_0,b)\in [3,4]\times[2,3]$ in Fig.~\ref{fig:quasi_diagram}.
For this case, we have
\begin{align}\label{eq:cevol2}
	&\vec{c}\,(t=\tACn{1}^+)\notag\\
	&=\mT_1 \mU_{1,2} \mT_2 (\mU_{2,3})^2 \mT_2\mU_{1,2}\mT_1\vec{c}\,(t=-\tACn{1}^-),
\end{align}
with which Eq.~\eqref{eq:PupC3} gives $\Pup$.
Even for this case, the FLZ theory~\eqref{eq:PupC3} well reproduces $\Pup$ as shown in Fig.~\ref{fig:PLZ} away from the narrow region near $a_0=\aACn{2}=3.05$.
The discrepancy in this narrow region is due to the adiabatic-impulse approximation.

\begin{figure}
	\includegraphics[width=\columnwidth]{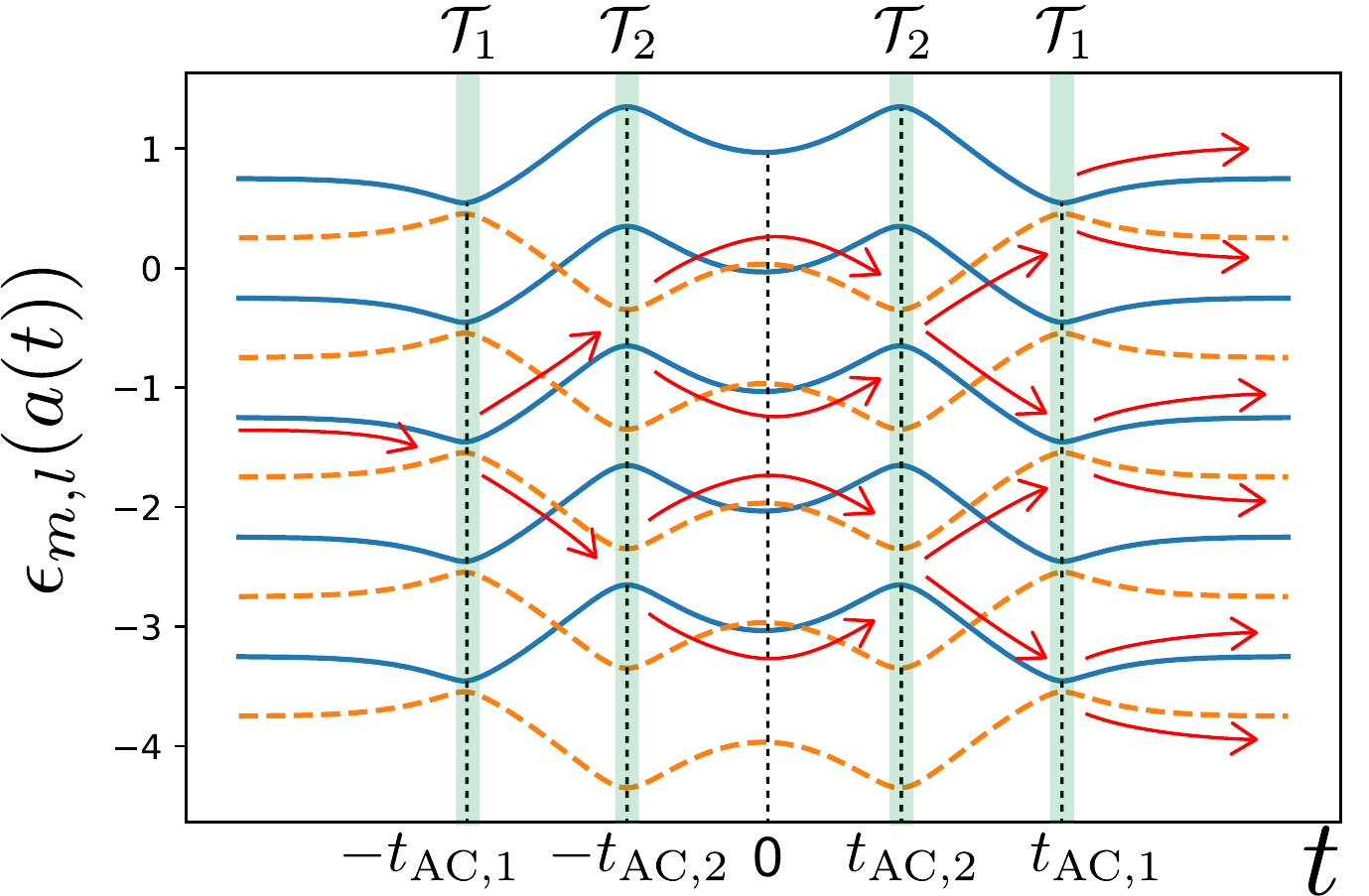}
	\caption{Schematic illustration of the Floquet-Landau-Zener interferometry for $\aACn{2}<a_0<\aACn{3}$.
	Solid (dashed) lines show those for Floquet states of $m=2$ ($m=1$) approaching $\ket{\downarrow}$ ($\ket{\uparrow}$) as $a\to0$.
	The curved arrows show quantum state trajectories along the IFS,
	and $\mT_1$ ($\mT_2$) denotes the transfer matrix for the avoided crossing at $\aACn{1}$ ($\aACn{2}$).
	}
	\label{fig:quasi_diagram2}
\end{figure}

With the transfer matrix formulation, we can finally interpret the complex pattern in $(a_0,b)\in [3,4]\times[2,3]$ in Fig.~\ref{fig:mask}.
In this parameter region, there are four passages of avoided crossings, under which the state flow of Eq.~\eqref{eq:cevol2} is schematically illustrated in Fig.~\ref{fig:quasi_diagram2}.
After the strong pulse irradiation, three Floquet states are superposed for each $m=1$ and $2$.
Thus, we have more quantum-path interference than other parameter regions like the previous case~\eqref{eq:PupN1}.
The complex pattern in $\Pup$ is understood qualitatively and quantitatively by the Landau-Zener-St\"{u}ckelberg interferometry in terms of the IFS.

%###

%#######################
\section{Pulse-width dependence}\label{sec:width}
In Sec.~\ref{sec:numerical}, we fixed the pulse width as $\nu=6$.
Meanwhile, experimentally, stronger peak amplitudes $a_0$ tend to be realized for shorter pulse widths~\cite{Brabec2000}.
Thus, it is crucially important how small $\nu$ can be for the FLZ theory remaining applicable.
Naively speaking, the FLZ theory is expected to become worse for shorter pulses because the envelope's temporal change $da/dt$ increases, and the assumption of adiabaticity eventually breaks down (see also Eq.~\eqref{eq:FHam}).
Note that this tendency also holds for increasing amplitude $a_0$ as $da/dt$ increases with $a_0$ as well.

\begin{figure}
	\includegraphics[width=\columnwidth]{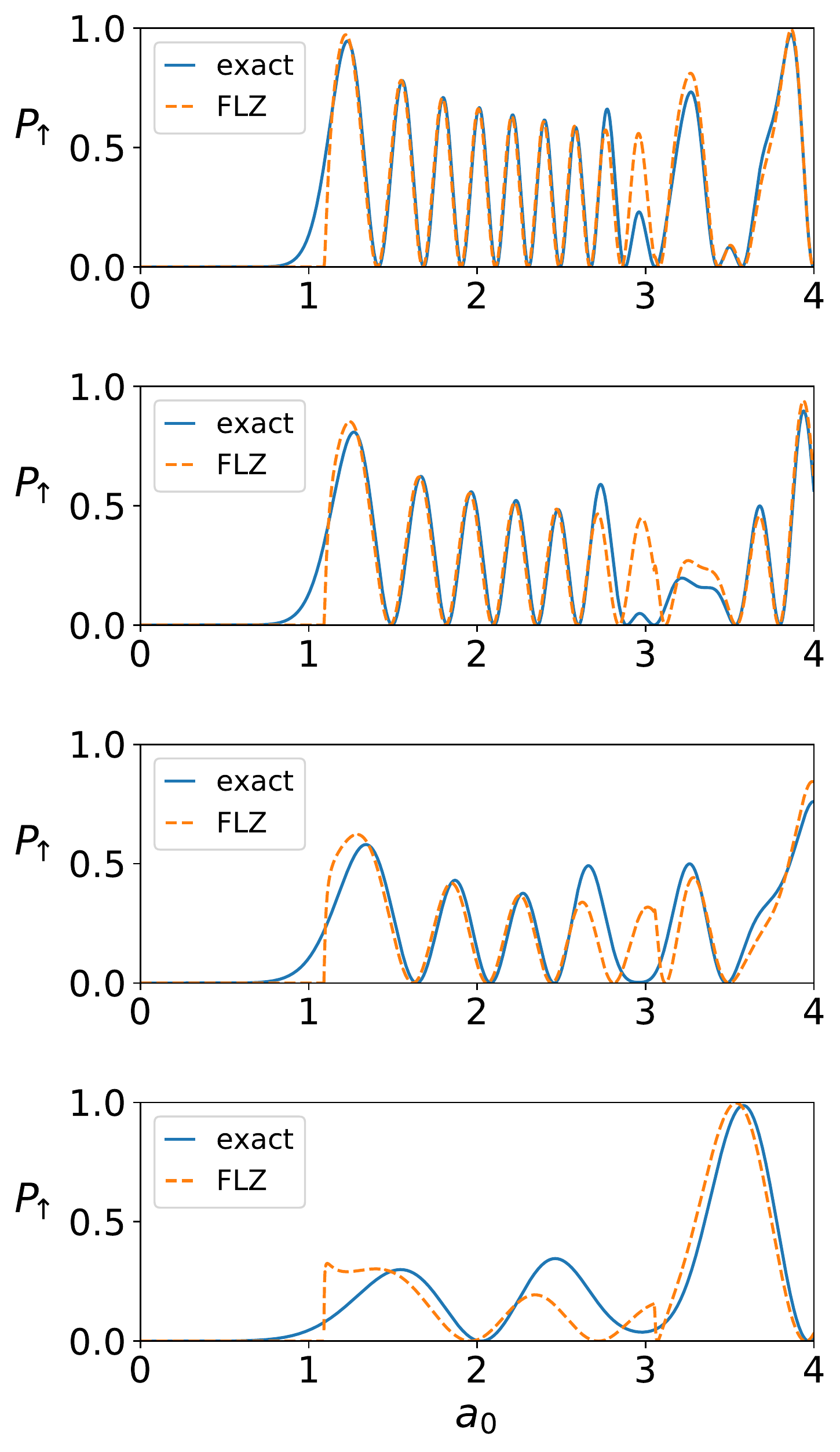}
	\caption{Excitation probability $P_\uparrow$ for $b=2.5$ plotted against the pulse peak height $a_0$. The solid and dashed curves show $P_\uparrow$ obtained, respectively, by solving the TDSE~\eqref{eq:Schpulse} numerically and by invoking the FLZ theory~\eqref{eq:PupC3}.
	The pulse width is $\nu=4,3,2,$ and $1$ from top to bottom.}
	\label{fig:widthdep}
\end{figure}

To address this issue of pulse-width dependence, we examine the FLZ theory's applicability with decreasing $\nu$.
Figure~\ref{fig:widthdep} shows $\Pup$ for $b=2.5$ obtained by the exact numerical simulation of the TDSE~\eqref{eq:Schpulse} and by the FLZ method for different $\nu$'s.
For $\nu=2,3$, and 4, we observe that the FLZ theory captures quite well the exact results within the adiabatic-impulse approximation except for the regions near the avoided crossing points $a=\aACn{1}=1.09$ and $\aACn{2}=3.05$.

We note that these regions of disagreement become wider for smaller $\nu$, which is consistent with the following intuitive argument on adiabaticity.
Our FLZ theory within the adiabatic-impulse approximation assumes that no transition occurs between IFSs except $a=\aACn{n}$ $(n=1,2,\dots)$.
However, this assumption is valid when the quasienergy difference is much larger than the perturbation term ($\mG$ in Eq.~\eqref{eq:FHam}) proportional to $da/dt$.
Thus, this assumption is not satisfied near the avoided crossings where the quasienergy differences become small.
Also, this tendency is stronger for shorter pulses having larger $da/dt$.
Transitions between IFSs actually occur slightly away from the avoided crossings, while the adiabatic-impulse approximation neglects them entirely.
Nevertheless, it is remarkable that the FLZ theory works quantitatively well and has wide-enough applicability parameter regimes even if the pulse is as short as 2-cycle ($\nu=2$).

For the single-cycle pulse ($\nu=1$), the FLZ theory does not agree well with the numerically exact results as seen in the range of $\aACn{1}<a_0<\aACn{2}$, indicating that the FLZ theory does not provide a good physical interpretation.
There are two possible reasons for the breakdown of the FLZ theory in $\nu<2$:
(i) the adiabatic impulse approximation becomes inaccurate, and
(ii) transitions to distant Floquet replicas become nonnegligible.
As for (i), we notice that $\adi\equiv1/\nu$ serves as the adiabatic parameter (see Appendix~\ref{app:semiclassical} for details).
Since the adiabatic impulse approximation is justified in two-level systems by power-series arguments for $\adi$ and is accurate for small $\adi$~\cite{Drese1999,Berry1990}, it is natural that the approximation here starts to fail at $\adi\sim1$, i.e., $\nu\sim1$, although their precise values should depend on models.
The other possibility (ii) is unlikely because the transition matrix elements $\mG_{\alpha\beta}(a)$ between IFSs are actually small compared to the quasienergy difference $\epsilon_\alpha(a)-\epsilon_\beta(a)$ (see Appendix~\ref{app:transitions} for details).
Thus, the possibility (i) is likely to be the reason for the FLZ's failure at the ultrashort pulse as short as $\nu\sim1$ in our model.

Despite this argument on the FLZ theory's failure for ultrashort pulses, we leave for future work to determine the precise value of $\nu$ for the breakdown and to fully understand why this theory works even down to $\nu=2$.
It is worth noting that the FLZ theory seems to work for $a_0>\aACn{2}$ at $\nu=1$, even with the disagreement in $a_0<\aACn{2}$, for which the authors have not found its reason.
Systematically studying the FLZ in such an ultrashort-pulse regime is an open problem.

\section{Summary and Discussions}\label{sec:summary}
Considering the pulse excitation probability $\Pup$ in a two-level quantum system, we have studied the complex interference pattern in Fig.~\ref{fig:mask} in the two-dimensional space spanned by the pulse peak height $a_0$ and the two levels' energy difference $b$.
To understand these patterns, we have utilized the instantaneous Floquet states (IFS), rather than the original energy eigenstates, as a useful basis for understanding dynamics~\cite{Breuer1989,Drese1999}.
The time evolution driven by strong pulse fields can then be regarded as adiabatic evolutions along the IFS and Landau-Zener-type (LZ-type) diabatic transitions between them~\cite{Holthaus2015}.

We have developed this idea quantitatively by applying the transfer matrix method among Floquet states, formulated how to keep track of quantum states under multiple LZ-type transitions, and termed this formulation as the Floquet-Landau-Zener (FLZ) theory in Sec.~\ref{sec:theory}.
Implementing this theory numerically in Sec.~\ref{sec:numerical} (and analytically in Appendix.~\ref{sec:analytical}), we have shown that the FLZ theory well reproduces $\Pup$ obtained by direct numerical calculations.
One advantage of the FLZ theory is that the physical interpretation of dynamics is transparent; The complex interference patterns in $\Pup$ originate from quantum path interference between IFSs as illustrated in Figs.~\ref{fig:quasi_diagram} and \ref{fig:quasi_diagram2}.

We have demonstrated that the FLZ theory is valid for longer pulses (i.e., larger $\nu$).
This is a natural tendency because the longer pulses mean slower changes of pulse envelopes $a(t)$, validating the adiabatic approximation.
Rather surprisingly, however, the FLZ theory has worked in appropriate parameter ranges if the pulse width is larger roughly than 2 cycles ($\nu\gtrsim2$) as shown in Sec.~\ref{sec:width}.
This should be relevant for experimental studies to address Floquet-related physics, which emerge ideally under strong continuous external fields while strong laser fields experimentally tend to be realized in short pulses.
Our findings imply that Floquet-related physics are present even in short-pulse experiments if we interpret appropriately in the sense of IFSs.
For the extension of the Floquet formalism to the case not strictly time-periodic external fields,
one of the present authors and Mizumoto~\cite{Kayanuma2000} studied the transition dynamics in the two-level system
under the level-crossing with a constant velocity plus time-periodic modulation in the relative energy.
An unexpected agreement has been observed between the calculation by the Floquet-Landau-Zener transfer matrix
method and the numerical solutions of the TDSE for a wide range of parameter values.
A deep understanding of this success is also an open question.

%####
As concluding remarks, we list some future directions.
First, it is important to validate the transfer matrix methods in the Floquet extended space and to improve the adiabatic-impulse approximation systematically.
For the conventional Landau-Zener problem, the so-called Stokes phenomena are known to underlie~\cite{Davis1976}, and WKB theories~\cite{Taya2021a} provide mathematical foundations.
One could generalize these insights to the Floquet extended space and validate the FLZ theory mathematically.
Second, it is intriguing to realize the FLZ interferometry in an experiment.
Our model should apply to any two-level systems, but there are two experimental challenges: (i) a long-enough coherence time and (ii) a strong-enough coupling to external fields with long-enough pulses.
In the present work, we neglected any decoherence/dissipation, which, if strong, may destroy the clear interference pattern.
It is intriguing to investigate if we can overcome those effects experimentally together with further theoretical investigations.
Finally, it is interesting to generalize the FLZ interferometry for other classes of systems with more than two levels~\cite{Holthaus1994a}, including multiple two-level systems~\cite{Niranjan2020}.
Generally speaking, the denser the energy levels are, the worse the adiabatic approximation becomes.
Thus, we expect that the FLZ theory works in systems with not-so-many levels.
However, it could be possible to apply this theory for condensed-matter systems with many energy levels but an energy gap above the ground state.
We leave the above problems open for future studies.

\section*{Acknowledgements}
T.N.I. thanks M. Holthaus for introducing the IFS formalism and M. Hongo and H. Taya for fruitful discussions on related topics.
Y.K. thanks Professors K. Yamanouchi and E. L\"{o}tstedt for drawing his attention to the works of population inversion in N${}_2$ molecules under irradiation of intense pulse laser~\cite{Zhang2017a}.
This work was supported by JSPS KAKENHI Grant No. JP21K13852 and JP19K03696,
and by Non-Equilibrium Working group (NEW) at RIKEN Interdisciplinary Theoretical and Mathematical Sciences Program (iTHEMS).

\appendix

%######
\section{Transfer matrix in the Landau-Zener problem}\label{sec:LZ}
\begin{figure}
	\includegraphics[width=\columnwidth]{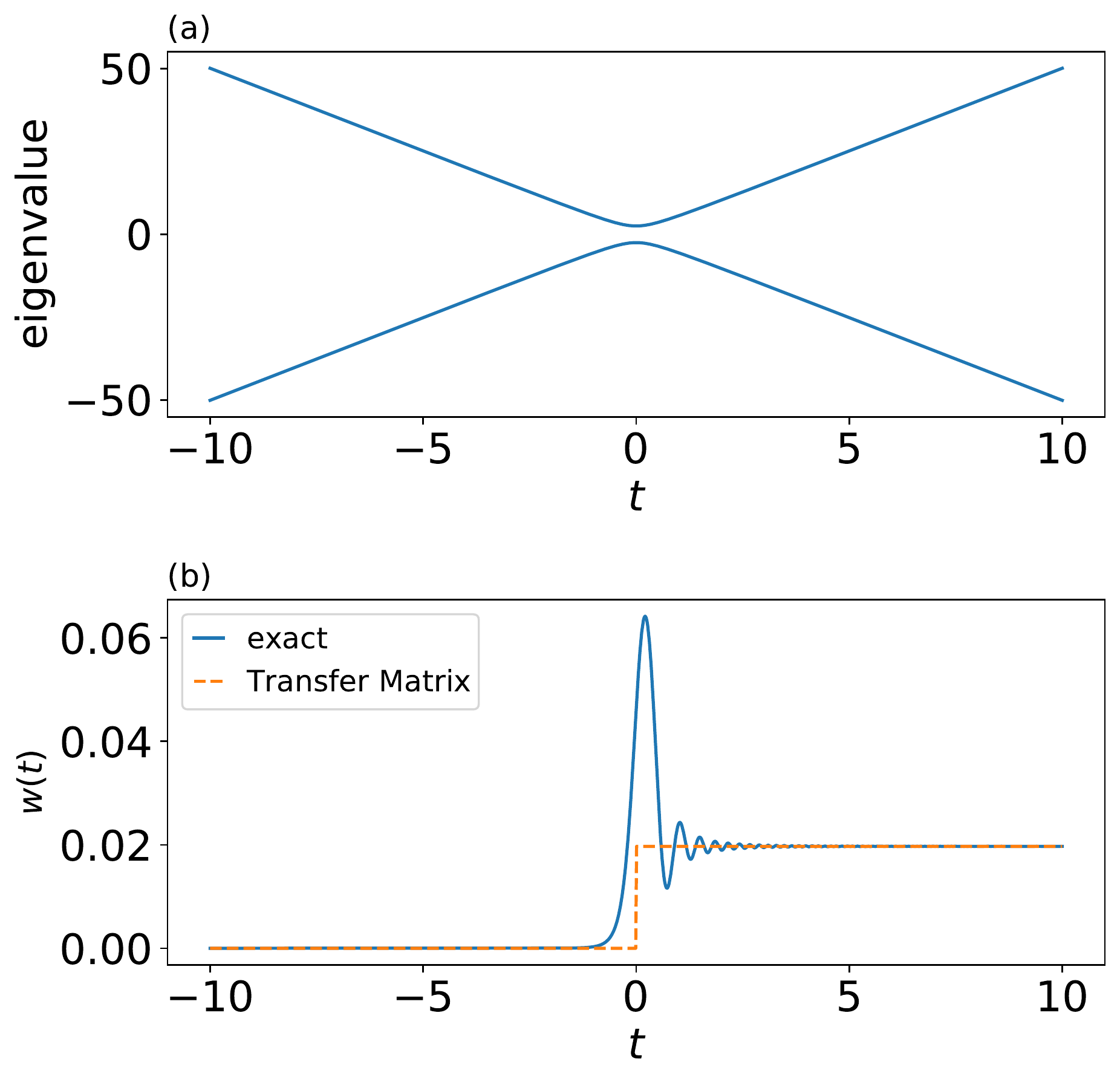}
	\caption{(a) Eigenvalues of $\HLZ(t)$ (see also text). (b) Upper-state populations $w(t)$ calculated by the exact numerics (solid) and by the transfer matrix method within the adiabatic-impulse approximation (dashed). In both panels, we set $v=10$ and $\Delta=5$.} 
	\label{fig:LZdemo}
\end{figure}

To supplement the discussions of transfer matrices for IFSs in the main text, we briefly review the transfer matrix for the seminal Landau-Zener problem for the linearly-time-dependent Hamiltonian $\HLZ(t)=-(vt/2)\sigma_z+(\Delta/2)\sigma_x=\sum_{n=1}^2 E_n(t)\ket{E_n(t)}\bra{E_n(t)}$ with eigenenergies $E_1(t)\ge E_2(t)$.
As shown in the level diagram in Fig.~\ref{fig:LZdemo}(a), from $t=-\infty$ to $+\infty$, the system goes across, at $t=0$, the avoided crossing of gap $\Delta=E_1(0)-E_2(0)$ at speed $v$.

Suppose that the initial state is given as $\ket{\Phi(\tin)}$ ($\tin<0$) and we are to solve its evolution $\ket{\Phi(t)}$ and ask the population at the upper state $w(t)=|\braket{E_1(t)|\Phi(t)}|^2$.
This population can be obtained analytically~\cite{Berry1990,Shevchenko2010} or numerically as illustrated in Fig.~\ref{fig:LZdemo}(b), where $\Delta=5$, $v=10$, $\tin=-10$, and $\ket{\Phi(\tin)}=\ket{E_2(\tin)}$.

%#######

%####
The transfer matrix method enables us to obtain an approximate solution with clear physical interpretation.
This method is based on the energy eigenbasis $\ket{\Phi(t)}=\sum_{n=1}^2 b_n(\tin)\ket{E_n(\tin)}$ and the assumption that no transition occurs away from the avoided crossing point $t=0$.
This assumption is known as the adiabatic-impulse approximation~\cite{Shevchenko2010}.
With this method, we have, for $n=1$ and $2$, $b_n(t<0)=\exp[-i\int_{\tin}^t ds E_n(s)]b_n(\tin)$ and $b_n(t>0)=\exp[-i\int_{\tin}^t ds E_n(s)]b_n(0+)$.
At the avoided crossing, the state experiences transitions described by the following transfer matrix,
\begin{align}
	\bm{b}(0+) =
	\begin{pmatrix}
		\sqrt{1-P}e^{-i\varphi^S} & -\sqrt{P} \\
		\sqrt{P} & \sqrt{1-P} e^{i\varphi^S}
	\end{pmatrix}
	\bm{b}(0-),
\end{align}
where 
\begin{align}
	P \equiv \exp(-2\pi \delta)
\end{align}
is the prominent Landau-Zener transition probability with
\begin{align}
	\delta &\equiv \frac{\Delta^2}{4v},
\end{align}
and 
\begin{align}
	\varphi^S &\equiv -\frac{\pi}{4}+\delta \ln(\delta-1) + \arg\Gamma(1-i\delta),
\end{align}
is the Stokes phase with $\Gamma(z)$ being the gamma function.
As illustrated in Fig.~\ref{fig:LZdemo}(b), the transfer matrix method well captures the exact solution except in the vicinity of the avoided crossing.
While the actual dynamics is complicated within the avoided crossing region, it well describes, as a single matrix, the integrated evolution from the input to the output.

%####
The transfer matrix method applies not only to $\HLZ(t)$ of linear time dependence but also to other Hamiltonians of more generic dependence~\cite{Shevchenko2010}.
For generic cases, we define $v$ and $\Delta$ by linearly approximating the energy eigenvalues near the avoided crossing (see, e.g., Ref.~\cite{Kayanuma2000}).
This generality enables us to apply this method for the IFS of our interest.

%######################
\section{Derivation of Eq.~\eqref{eq:IFSevol}}\label{sec:c_eq}
Here we derive Eq.~\eqref{eq:IFSevol} by substituting Eq.~\eqref{eq:IFSexpansion} into the Schr\"{o}dinger equation $id\ket{\Psi(t)}/dt = \Hpulse (t)\ket{\Psi(t)}$.
Note that $\Hpulse (t)=\Hcw(a(t),t)$ by definition, and the Schr\"{o}dinger equation reads
\begin{align}\label{seq:Sch}
	i\frac{d\ket{\Psi(t)}}{dt} = \Hcw(a(t),t)\ket{\Psi(t)}.
\end{align}
The left-hand of Eq.~\eqref{seq:Sch} side becomes
\begin{widetext}
\begin{align}\label{seq:dPsidt}
i\frac{d\ket{\Psi(t)}}{dt} =
	i\sum_\alpha \left[ \frac{dc_\alpha(t)}{dt} \ket{u_\alpha(a(t),t)} + c_\alpha(t)\frac{da}{dt}\frac{\partial \ket{u_\alpha(a(t),t)}}{\partial a}  + c_\alpha(t)\frac{\partial \ket{u_\alpha(a(t),t)}}{\partial t} \right].
\end{align}
\end{widetext}

Here we recall the definition of Floquet states,
$i\frac{\partial}{\partial t} [e^{-i\epsilon_\alpha(a(t)) t}\ket{u_\alpha(a(t),t)} ] = \Hcw (a(t),t)[e^{-i\epsilon_\alpha(a(t)) t}\ket{u_\alpha(a(t),t)} ]$,
which implies
\begin{align}\label{seq:dudt}
i\frac{\partial \ket{u_\alpha(a(t),t)}}{\partial t} = [-\epsilon_\alpha(a(t)) + \Hcw (a(t),t)]\ket{u_\alpha(a(t),t)}.
\end{align}
Then we substitute Eq.~\eqref{seq:dPsidt} together with Eq.~\eqref{seq:dudt} into Eq.~\eqref{seq:Sch}, obtaining
\begin{align}
	\sum_\alpha \left[i\frac{dc_\alpha(t)}{dt} -\epsilon_\alpha(a(t))\right] \ket{u_\alpha(a(t),t)}\notag\\
	+ i\sum_\alpha c_\alpha(t)\frac{da}{dt}\frac{\partial \ket{u_\alpha(a(t),t)}}{\partial a}
	=0,\label{seq:ceq1}
\end{align}
where the terms with $\Hcw$ canceled out between the left- and right-hand sides.
%Finally, we left-multiply $\bra{U_\beta(a(t),t)}$ onto Eq.~\eqref{seq:ceq1} and use the orthonormality $\int_0^T(dt/T)\braket{u_\beta(a(t),t)}$

Finally, to rewrite the second term of Eq.~\eqref{seq:ceq1}, we note the following relations
\begin{align}
	&\frac{\partial \ket{u_\alpha(a(t),t)}}{\partial a}\notag\\
	&= \int_0^T dt' \delta_T(t-t') \frac{\partial \ket{u_\alpha(a(t),t')}}{\partial a}\notag\\
	&= \sum_\beta \int_0^T \frac{dt'}{T} \ket{u_\beta(a(t),t)}\bra{u_\beta(a(t),t')} \frac{\partial \ket{u_\alpha(a(t),t')}}{\partial a}\notag\\
	&=\sum_\beta \ket{u_\beta(a(t),t)}\mathcal{G}_{\beta\alpha}(a(t)),\label{seq:useG}
\end{align}
where we used Eqs.~\eqref{eq:identity} and \eqref{eq:defG} to obtain the third and fourth lines, respectively.
Substituting Eq.~\eqref{seq:useG} into Eq.~\eqref{seq:ceq1} and considering the coefficients of each $\ket{u_\alpha(a(t),t)}$, we obtain
\begin{align}
	i\frac{dc_\alpha(t)}{dt} = \epsilon_\alpha(a(t)) -i \frac{da}{dt}\sum_\beta \mathcal{G}_{\alpha\beta}(a(t)) c_\beta(t),
\end{align}
which is equivalent to Eq.~\eqref{eq:IFSevol} in the matrix representation.

%#######################
\section{Analytical approach to circular and elliptic polarizations}\label{sec:analytical}
In this appendix, we consider the following coupling term
\begin{align}\label{eq:Vt_ell}
	V(t) = \frac{1+\lambda}{2}\cos(\omega t) \sigma_x
    +\frac{1-\lambda}{2}\sin(\omega t) \sigma_y,
\end{align}
which reduces to Eq.~\eqref{eq:Vt} for $\lambda=1$.
For a single spin-1/2, this term represents the Zeeman coupling $V(t)=\bm{B}(t)\cdot \bm{\sigma}$ to an elliptically-polarized magnetic field $\bm{B}(t)=(\frac{1+\lambda}{2}\cos(\omega t),\frac{1-\lambda}{2}\sin(\omega t),0)$.
The dimensionless parameter $\lambda$ quantifies the ellipticity, and the special values $\lambda=0$ and $1$ correspond to the circular and linear polarizations, respectively.
Therefore, we call $\lambda=0$, 1, and the others as the linear, circular, and elliptic polarizations, respectively, even if the model does not necessarily suppose a single spin-1/2.

In the main text, we have shown that the FLZ theory works well for the linear polarization $(\lambda=1)$.
In those calculations, we implemented the transfer matrices $\mT_n$ and phase acquisition operators $\mU_{n+1,n}$ constructed from the quasienergies $\epsilon_m(a)$ obtained numerically.
When $\lambda$ is zero or small, however, we can analytically obtain the quasienergies approximately using the perturbation theory for $\lambda$.
In this appendix, using this analytical approach, we extend analyses and gain deeper insights from the limit of $\lambda=0$ to small $\lambda$.

\subsection{Circular polarization}\label{sec:circ}
We begin by considering the circular polarization ($\lambda=0$), for which the coupling term reads
\begin{align}
	V_0(t) = e^{-i\omega t}\sigma_+ + e^{i\omega t} \sigma_- \label{eq:Vrot}
\end{align}
with $\sigma_\pm=(\sigma_x\pm i\sigma_y)/2$.
Note that this case corresponds to the rotating-wave approximation of Eq.~\eqref{eq:Vt}.
In this special case, the continuous-wave problem~\eqref{eq:Schcw} corresponds to the seminal Rabi model~\cite{Shirley1965}.
We analytically obtain the two independent solutions as
\begin{align}
	\ket{\psi_A(t)} &= \frac{e^{-i(\Omega+\frac{\omega}{2})t}}{\sqrt{2\Omega[\Omega+(\omega-b)/2]}}
	\begin{pmatrix}
	a/2 \\ \left(\frac{\omega-b}{2}+\Omega \right)e^{i\omega t}		
	\end{pmatrix}, \label{eq:psiA}\\
		\ket{\psi_B(t)} &= \frac{e^{i(\Omega+\frac{\omega}{2})t}}{\sqrt{2\Omega[\Omega-(\omega-b)/2]}}
	\begin{pmatrix}
	\frac{a}{2}e^{-i\omega t} \\ \frac{\omega-b}{2}-\Omega 
	\end{pmatrix},\label{eq:psiB}
\end{align}
where
\begin{align}
	\Omega = \frac{1}{2}\sqrt{a^2 + (b-\omega)^2}
\end{align}
is the Rabi frequency (energy).

\begin{figure}
	\begin{center}
	\includegraphics[width=\columnwidth]{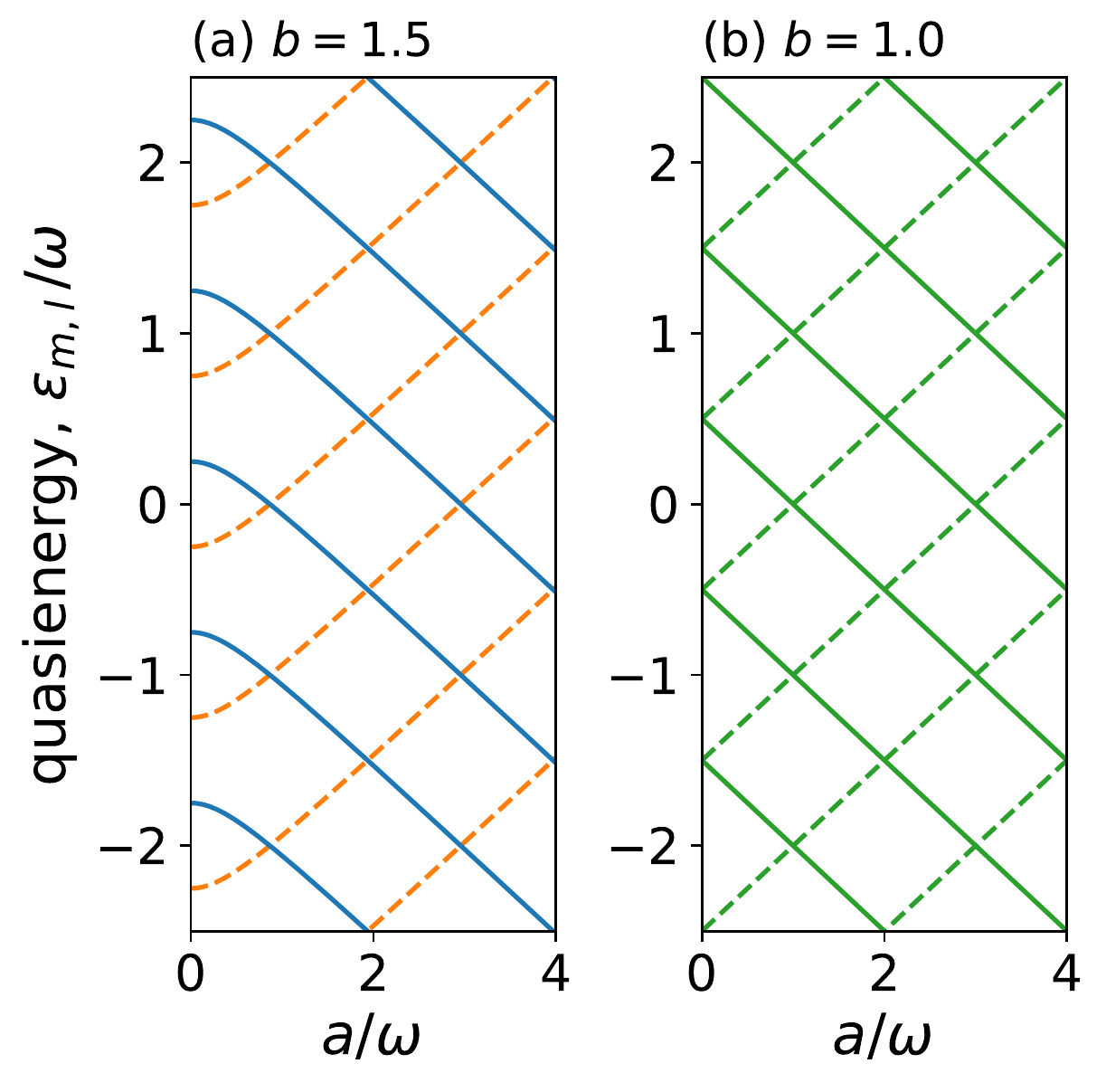}
	\caption{Quasienergies for $\lambda=0$ in (a) $b=1.5$ and (b) $b=1.0$ plotted against coupling strength $a$.
	(a) Solid (dashed) lines show those
	for Floquet states of $m=2$ ($m=1$) approaching $\ket{\downarrow}$ ($\ket{\uparrow}$) as $a\to0$.
	(b) Solid and dashed lines show those
	for Floquet states corresponding to $\ket{\psi_B(t)}$ and $\ket{\psi_A(t)}$ approaching $(\ket{\uparrow}\mp\ket{\downarrow})/\sqrt{2}$ as Eqs.~\eqref{eq:lim1} and \eqref{eq:lim2}, respectively.
	}
	\label{fig:quasi_circ}
	\end{center}	
\end{figure}

%#####
We can read out the Floquet states and their quasienergies from Eqs.~\eqref{eq:psiA} and \eqref{eq:psiB}.
Recall that, in this work, we assign the Floquet states' label $(m,l)$ so that $(m,0)$ approaches the undriven solutions $e^{-i(b/2)t}\ket{\uparrow}$ ($e^{+i(b/2)t}\ket{\downarrow}$) for $m=1$ ($m=2$).
To make this assignments, it is convenient to consider the two cases, $b>\omega$ and $b<\omega$, separately.
For $b>\omega$,
\begin{align}
\ket{\psi_A(t)}\to  e^{-i(b/2)t}\ket{\uparrow},\ %\begin{pmatrix} 1 \\ 0 \end{pmatrix},
\ket{\psi_B(t)}\to  e^{+i(b/2)t}\ket{\downarrow} %\begin{pmatrix} 0 \\ 1 \end{pmatrix}
\end{align}
in the limit of $a\to0$.
Thus, in this case, we see that 
\begin{align}
	\ket{u_1(a;t)} &= \frac{1}{\sqrt{2\Omega[\Omega+(\omega-b)/2]}}
	\begin{pmatrix}
	a/2 \\ \left(\frac{\omega-b}{2}+\Omega \right)e^{i\omega t}		
	\end{pmatrix}, \label{eq:u1-a}\\
		\ket{u_2(a;t)} &= \frac{1}{\sqrt{2\Omega[\Omega-(\omega-b)/2]}}
	\begin{pmatrix}
	\frac{a}{2}e^{-i\omega t} \\ \frac{\omega-b}{2}-\Omega 
	\end{pmatrix},\label{eq:u2-a}
\end{align}
and all the quasienergies~\eqref{eq:quasi_replica} are 
\begin{align}
\epsilon_{1,l}(a) &=+\Omega+\frac{\omega}{2}+l\omega,\label{eq:quasi1-a}\\
\epsilon_{2,l}(a) &=-\Omega-\frac{\omega}{2}+l\omega.\label{eq:quasi2-a}	
\end{align}
We plot the quasienergy for a representative off-resonant case $b=1.5$ in Fig.~\ref{fig:quasi_circ}(a).
As the analytical expressions imply, there is no avoided crossing.

On the other hand, for $b<\omega$, 
\begin{align}
\ket{\psi_A(t)}\to  e^{+i(b/2)t} \ket{\downarrow},\ %\begin{pmatrix} 0 \\ 1 \end{pmatrix},\\
\ket{\psi_B(t)}\to  e^{i(\omega-b/2)t} \ket{\uparrow} %\begin{pmatrix} 1 \\ 0 \end{pmatrix}\\
\end{align}
as $a\to0$.
This means that
\begin{align}
	\ket{u_1(a;t)} &= \frac{e^{i\omega t}}{\sqrt{2\Omega[\Omega-(\omega-b)/2]}}
	\begin{pmatrix}
	\frac{a}{2}e^{-i\omega t} \\ \frac{\omega-b}{2}-\Omega 
	\end{pmatrix},\label{eq:u1-b}\\
	\ket{u_2(a;t)} &= \frac{1}{\sqrt{2\Omega[\Omega+(\omega-b)/2]}}
	\begin{pmatrix}
	a/2 \\ \left(\frac{\omega-b}{2}+\Omega \right)e^{i\omega t}		
	\end{pmatrix}, \label{eq:u2-b}
\end{align}
and all the quasienergies~\eqref{eq:quasi_replica} are 
\begin{align}
\epsilon_{1,l}(a) &=-\Omega+\frac{\omega}{2}+l\omega,\label{eq:quasi1-b}\\
\epsilon_{2,l}(a) &=+\Omega+\frac{\omega}{2}+l\omega.\label{eq:quasi2-b}	
\end{align}

%#####
On the resonance $b=\omega$, $\ket{\psi_A(t)}$ and $\ket{\psi_B(t)}$ do not converge to either $\ket{\uparrow}$ or $\ket{\downarrow}$ but to superpositions of them in the limit of $a\to0$.
In fact, we have, for $b=\omega$,
\begin{align}
\ket{\psi_A(t)} &\to  e^{-i(\omega/2)t} \frac{\ket{\uparrow}+\ket{\downarrow}}{\sqrt{2}},\label{eq:lim1} \\  %\begin{pmatrix} 0 \\ 1 \end{pmatrix},\\
\ket{\psi_B(t)} &\to  e^{+i(\omega/2)t} \frac{\ket{\uparrow}-\ket{\downarrow}}{\sqrt{2}} \label{eq:lim2} %\begin{pmatrix} 1 \\ 0 \end{pmatrix}\\
\end{align}
as $a\to0$.
We will see that this special property on the resonance $b=\omega$ leads to nontrivial behaviors of excitation probabilities $P_\uparrow$.
We plot the quasienergy for the resonant case $b=1$ in Fig.~\ref{fig:quasi_circ}(b), where they are degenerate at $a=0$.

%#####################

\begin{figure}
	\begin{center}
	\includegraphics[width=\columnwidth]{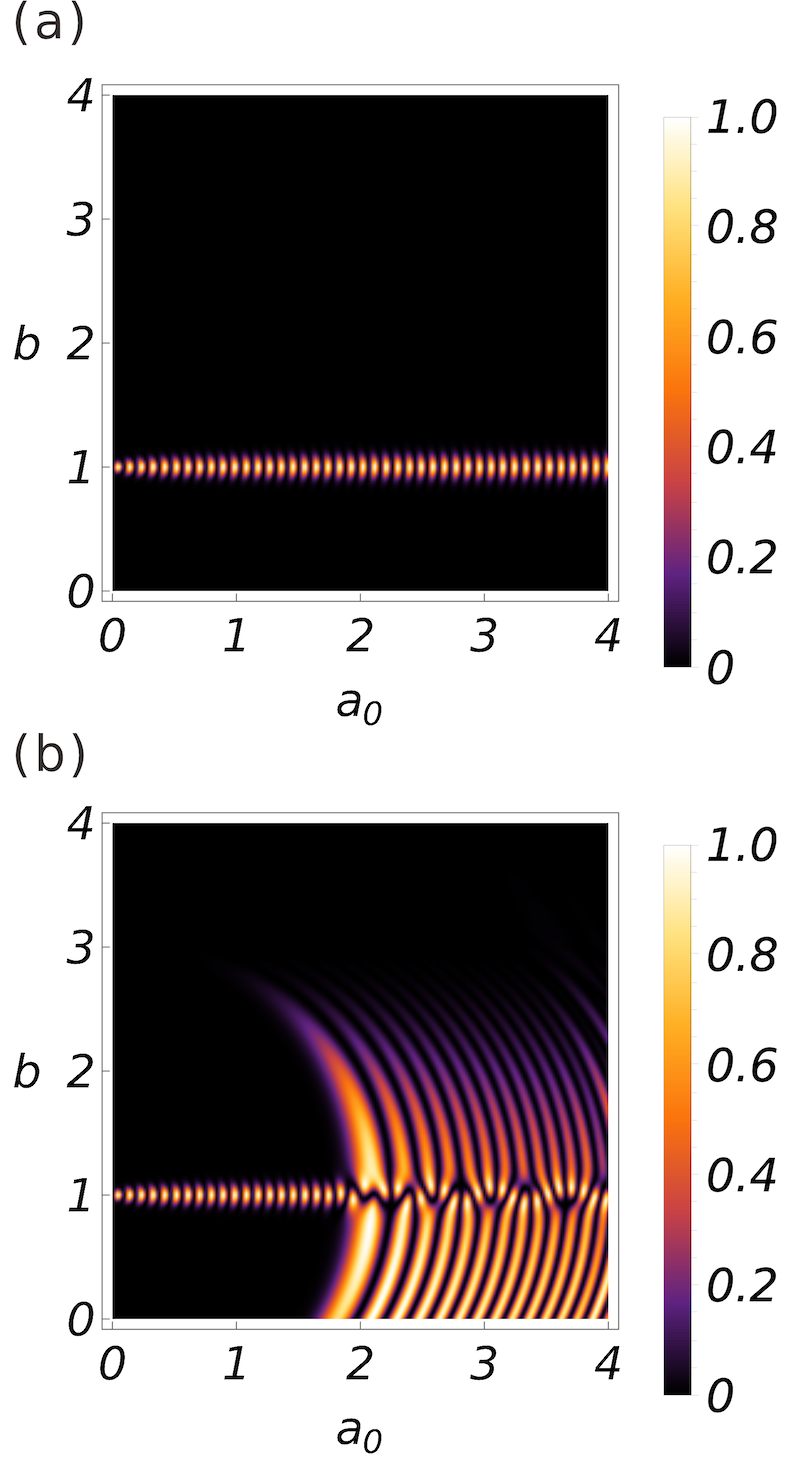}
	\caption{Excitation probability $P_\uparrow$ plotted against the pulse peak height $a_0$ and the energy level difference $b$ for (a) the circular ($\lambda=0$) and (b) an elliptic ($\lambda=0.1$) polarizations with pulse width $\nu=6$.}
	\label{fig:mask2}
	\end{center}	
\end{figure}

With these Floquet states, let us now study the pulse excitation problem for the circular polarization.
Figure~\ref{fig:mask2}(a) illustrates the excitation probability $\Pup$ for $\nu=6$, showing (i) almost no excitation away from the resonance $b\neq1$ and (ii) an oscillating behavior on resonance $b=1$.

%#####################

The property (i) is interpreted as follows.
Away from resonance, we have a quasienergy diagram like Fig.~\ref{fig:quasi_circ}(a) without avoided crossings, and each Floquet state approaches $\ket{\uparrow}$ or $\ket{\downarrow}$ as $a\to0$.
In such a case, the IFS interpretation goes like at the end of Sec.~\ref{sec:IFS}.
Our initial state for the pulse problem is $\ket{\downarrow}$, and it adiabatically moves along $\ket{u_2(a(t);t)}$, which coincides with $\ket{\downarrow}$ at $t=-\infty$.
In the adiabatic move, there occurs no LZ-type transition to other branches of Floquet states, and finally the state comes back to $\ket{\downarrow}$, meaning that $\Pup\approx0$.
Thus, the absence of avoided crossings, a special property of $\lambda=0$, explains the suppressed $\Pup$ for $b\neq1$ in Fig.~\ref{fig:mask2}(a).

The property (ii) is interpreted as follows.
On resonance, the Floquet states do not converge to either $\ket{\uparrow}$ or $\ket{\downarrow}$ but approach superpositions of them in $a\to0$.
Thus, our initial state $\ket{\downarrow}$ is a superposition of $\ket{u_1(a(t);t)}$ and $\ket{u_2(a(t);t)}$ at $t=-\infty$.
While these Floquet states move adiabatically without LZ-type transitions as there is no avoided crossing, they acquire relative phase factors due to the quasienergies.
Since the acquired phase from $t=-\infty$ to $t=+\infty$ is an increasing function of the pulse peak height $a_0$, the final excitation probability $\Pup$ oscillates with $a_0$.
This mechanism is FLZ interferometry discussed in Sec.~\ref{sec:numerical} although the superposition of Floquet states here is not created by LZ-type transitions but by particular limiting behaviors~\eqref{eq:lim1} and \eqref{eq:lim2}.

%############################
\subsection{Elliptic polarization}\label{sec:elliptic}
In Appendix~\ref{sec:circ}, we have shown that $\lambda=0$ is an ideal limit where the quasienergies are obtained analytically and no avoided crossing occurs.
For $\lambda\neq0$, $V(t)$ involves the counter-rotating component on top of Eq.~\eqref{eq:Vrot}:
\begin{align}
V(t) &= V_0(t) + \lambda W(t),\\
W(t)& =e^{i\omega t}\sigma_+ + e^{-i\omega t}\sigma_-.
\end{align}
This component hybridizes the independent solutions [Eqs.~\eqref{eq:psiA} and \eqref{eq:psiB}] for $\lambda=0$, giving rise to avoided crossings of quasienergies.
To address this scenario analytically, we here consider the case of small $\lambda$'s, i.e., nearly-circular elliptic polarizations.
In these cases, we can use the Floquet states for $\lambda=0$ (see Appendix~\ref{sec:circ}) as the unperturbed solutions and approximately obtain quasienergies with avoided crossings by perturbation theory in terms of $\lambda a_0$ (see, e.g., Ref.~\cite{Ikeda2018} for technicalities).

We remark that this approach does not assume $a_0$ is small.
In fact, the unperturbed solutions (Floquet states for $\lambda=0$) can involve nonperturbative effects of $a_0$.
Thus, this approach is particularly useful when
\begin{align}\label{eq:regime}
\lambda a_0 \ll 1 \ll a_0
\end{align}
since, for $a_0\ll1$, we can apply the naive perturbation theory in terms of $a_0$ for arbitrary $\lambda$.
We cannot find such parameters~\eqref{eq:regime} for the linear polarization ($\lambda=1$) that we studied in the main text, where we needed to calculate quasienergies numerically.

%###
%#####################
\begin{figure}
	\begin{center}
	\includegraphics[width=\columnwidth]{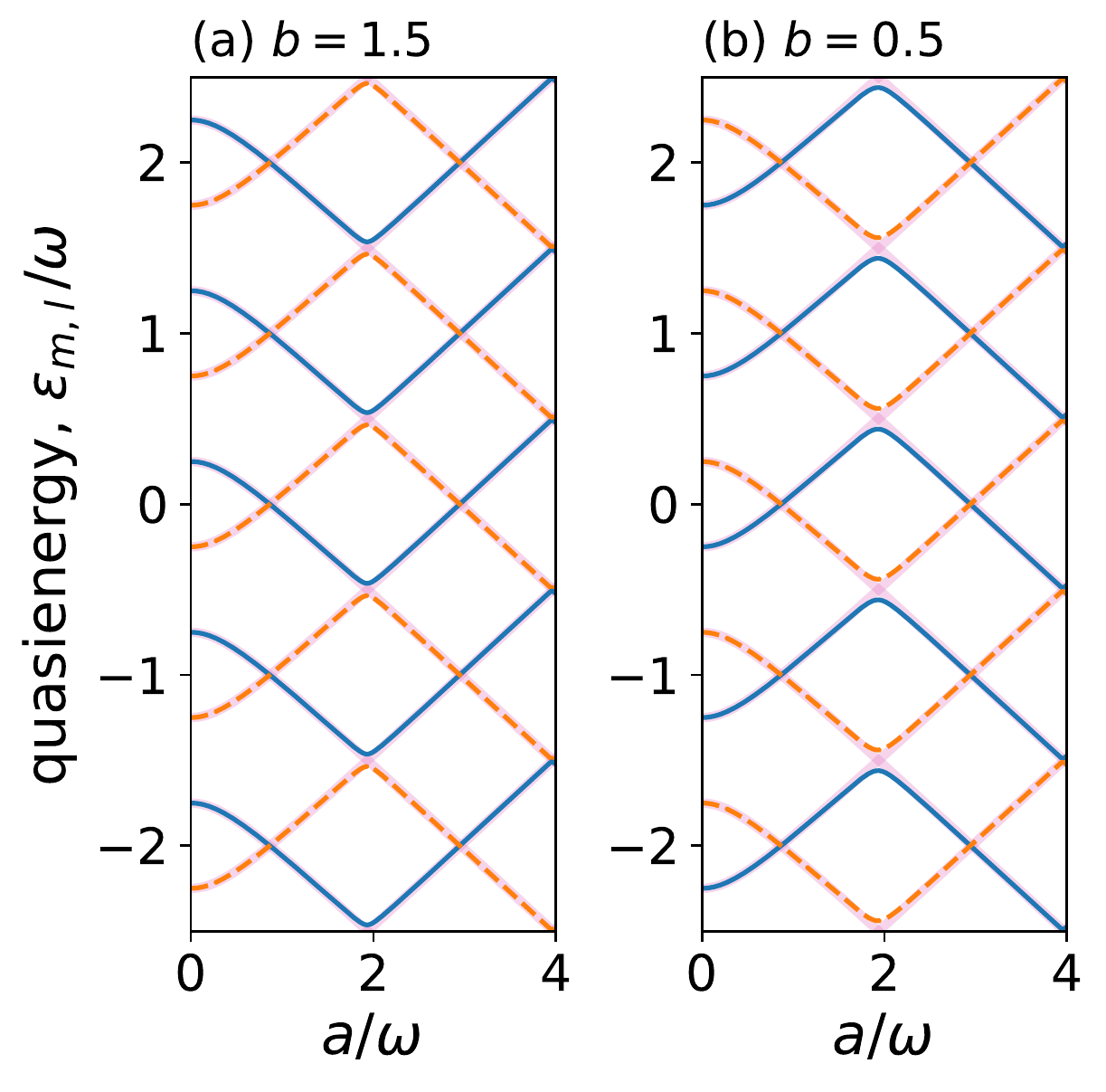}
	\caption{Quasienergies for $\lambda=0.1$ in (a) $b=1.5$ and (b) $b=0.5$ plotted against coupling strength $a$.
	Solid (dashed) lines show those
	for Floquet states of $m=2$ ($m=1$) approaching $\ket{\downarrow}$ ($\ket{\uparrow}$) as $a\to0$.
	Thicker background curves show quasienergies for $\lambda=0$ for reference.
	}
	\label{fig:quasi_elliptic}
	\end{center}	
\end{figure}

Before developing detailed calculations, we qualitatively see how the FLZ theory applies to elliptic polarizations.
Figure~\ref{fig:mask2}(b) illustrates the excitation probability $\Pup$ for $\lambda=0.1$ and $\nu=6$.
Unlike the circular polarization case [see panel (a)], $\Pup$ becomes significant away from the resonance condition $b=1$.
For, say, $b=1.5$ or $0.5$, as $a_0$ increases, $\Pup$ emerges at $a_0\approx2$ and then oscillates.
This behavior is consistent with the quasienergy diagram in Fig.~\ref{fig:quasi_elliptic}, which shows that the first avoided crossing appears at $\aACn{1}\approx2$.
Thus, the FLZ interpretation discussed in Sec.~\ref{sec:numerical} together with the quasienergy diagram qualitatively explains the interference pattern in Fig.~\ref{fig:mask2}(b).
We note the similarity between the quasienergies for $\lambda=0.1$ and $0$ shown in Fig.~\ref{fig:quasi_elliptic}; They are almost on top of each other away from the avoided crossings whereas they slightly repel each other at the avoided crossings.
This suggests the validity of considering $\lambda a_0$ as perturbation.

%####
Let us now quantitatively apply the FLZ theory developed in Sec.~\ref{sec:theory} with its inputs, such as $\aACn{n}$ and $\epsilon_{m,l}(a)$, obtained analytically.
For concreteness, we first focus on $1<b<3$, for which the quasienergy diagram looks like Fig.~\ref{fig:quasi_elliptic}(a).
As shown in the figure, although the quasienegies $\epsilon_{1,l}(a)$ and $\epsilon_{2,l+2}(a)$ overlap at $a\approx1$, they do not repel.
This is a selection rule due to the fact that $W(t)$ does not have matrix elements between these Floquet states,
\begin{align}
	\int_0^T \frac{dt}{T}\braket{u_{1,l}(a;t) | W(t) | u_{2,l+2}(a;t)}=0,
\end{align}
where $\ket{u_{m,l}(t)}=e^{il\omega t}\ket{u_m(t)}$ with Eqs.~\eqref{eq:u1-a} and \eqref{eq:u2-a}.
More generally, similar selection rules follow from the fact that the matrix elements vanish between $(1,l)$ and $(2,l+2k)$ ($k\in\mathbb{Z}$), which means physically that $2k$-photon transitions are prohibited.
Therefore, the first avoided crossing occurs at the 3-photon resonance defined by $\epsilon_{1,l}(a)=\epsilon_{2,l+3}(a)$, which gives
\begin{align}
	\aACn{1}=\sqrt{(\omega+b)(3\omega-b)} \qquad (\text{for}\ \omega<b<3\omega).\label{eq:aAC1_ell}
\end{align}
Note that we needed to calculate numerically $\aACn{1}$ for the linear polarization ($\lambda=1$) in Sec.~\ref{sec:numerical}.

%#####
The quasienergies in the presence of small $\lambda$ are obtained by considering the coupling by $\lambda W(t)$ between the unperturbed Floquet states $\alpha=(1,l)$ and $(2,l+3)$.
The Floquet Hamiltonian within the 2-dimensional subspace reads
\begin{align}
	\mathcal{H}_{\alpha\beta} =
	\begin{pmatrix}
		\epsilon_{1,l}(a) & \lambda K\\
		\lambda K & \epsilon_{2,l+3}(a)
	\end{pmatrix}_{\alpha\beta},\label{eq:22Hmat}
\end{align}
where $\alpha$ and $\beta$ denote either $(1,l)$ or $(2,l+3)$
and $\lambda K \equiv \int_0^T \frac{dt}{T}\braket{u_{1,l}(a) |\lambda W(t) | u_{2,l+3}(a)}$ yielding
\begin{align}
	K = \lambda \frac{a^2}{8\Omega}\sqrt{\frac{\Omega+\frac{\omega-b}{2}}{\Omega-\frac{\omega-b}{2}}}.
\end{align}
The eigenvalues of Eq.~\eqref{eq:22Hmat} lead to the quasienergies with avoided crossing:
\begin{align}
	\epsilon_{1,l}^{(\lambda)}(a;t) &\approx \frac{\omega}{2} - \sqrt{ \left(\frac{\epsilon_{1,l}(a)-\epsilon_{2,l+3}(a)}{2}\right)^2+(\lambda K)^2},\notag\\
	\epsilon_{2,l+3}^{(\lambda)}(a;t) &\approx \frac{\omega}{2} + \sqrt{ \left(\frac{\epsilon_{1,l}(a)-\epsilon_{2,l+3}(a)}{2}\right)^2+(\lambda K)^2}.\label{eq:eplambda}
\end{align}
In the approximation made here, we have ignored couplings outside the 2-dimensional subspace, and Eq.~\eqref{eq:eplambda} involves higher-order terms in $\lambda$.
The quasienergy gap at the avoided crossing follows from Eq.~\eqref{eq:eplambda} as
\begin{align}
	\Delta_1 &= \epsilon_{2,l+3}^{(\lambda)}(\aACn{1})-\epsilon_{1,l}^{(\lambda)}(\aACn{1})\notag\\
	%= \lambda \left. K \right|_{a=\aACn{1}}
	&=\lambda \frac{\sqrt{(\omega+b)(3\omega-b)^3}}{4\omega},\label{eq:gap_ell}
\end{align}
where we used $\epsilon_{1,l}(a)=\epsilon_{2,l+3}(a)$ and hence $\Omega=\omega$ at $a=\aACn{1}$.

\begin{figure}
	\includegraphics[width=\columnwidth]{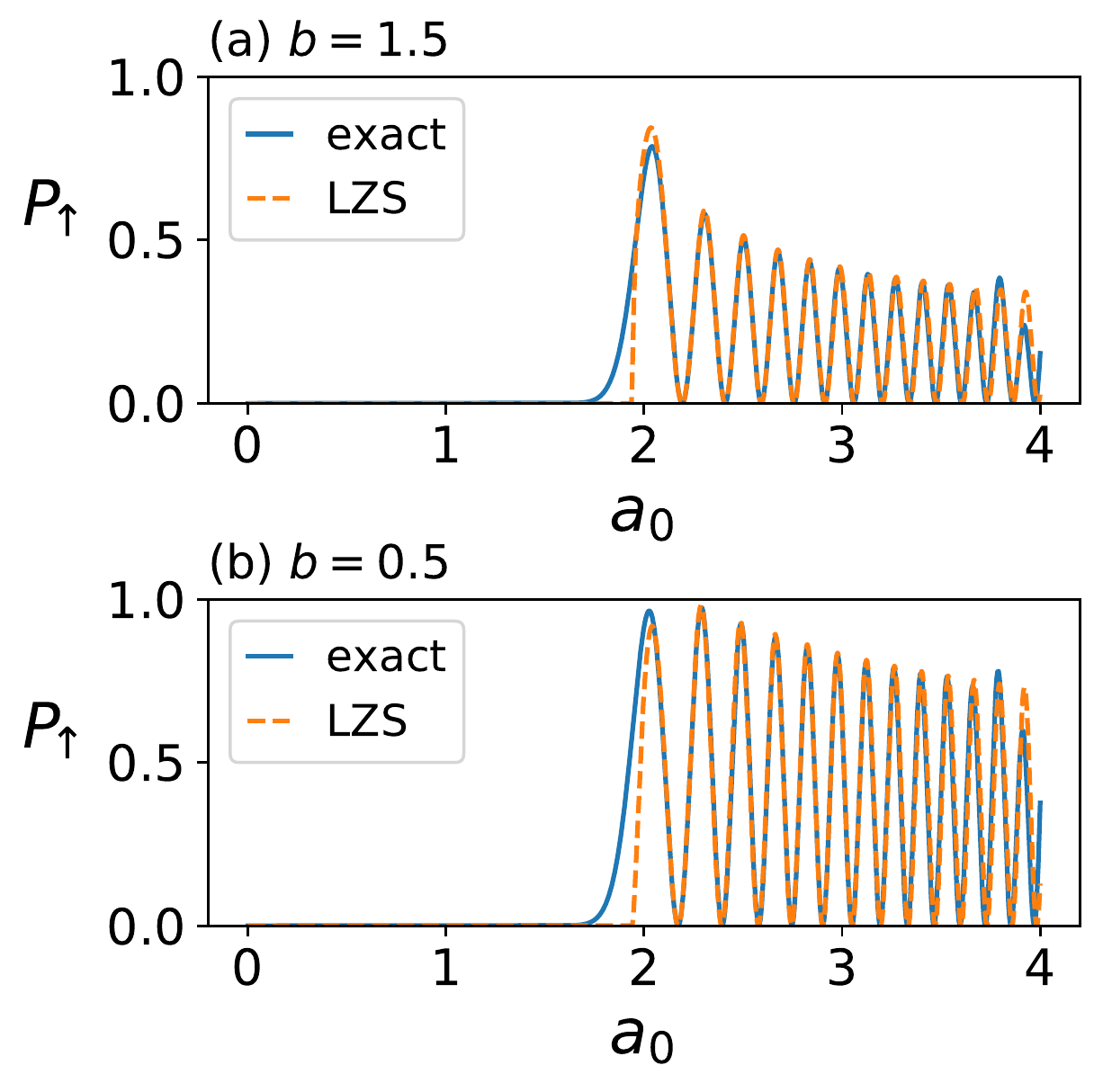}
	\caption{Excitation probability $P_\uparrow$ for (a) $b=1.5$ and (b) $b=0.5$ plotted against the pulse peak height $a_0$ for an elliptic polarization ($\lambda=0.1$) with pulse width $\nu=6$. The solid and dashed curves show $P_\uparrow$ obtained, respectively, by solving the TDSE~\eqref{eq:Schpulse} numerically and by invoking the FLZ theory~\eqref{eq:PupN1}.
	}
	\label{fig:PLZS_elliptic}
\end{figure}

%#####
Given that the avoided crossing point~\eqref{eq:aAC1_ell}, quasienergies~\eqref{eq:eplambda}, and their gap~\eqref{eq:gap_ell} are analytically obtained,
we can implement the FLZ theory quantitatively.
Restricting ourselves to $\aACn{1}<a_0<\aACn{2}$ for simplicity, we obtain $\Pup$ as in Eq.~\eqref{eq:PupN1},
where $P_1$ and $\varphi_1^S$ are obtained using Eqs.~\eqref{eq:Pn}, \eqref{eq:deltan}, and \eqref{eq:phisn},
and $\Phi_{m,l}$ are obtained combining Eqs.~\eqref{eq:Phi} and \eqref{eq:eplambda} with numerical evaluation of the integral.
The excitation probability $\Pup$ thus obtained is compared with the direct numerical solution for $b=1.5$, $\lambda=0.1$, and $\nu=6$ in Fig.~\ref{fig:PLZS_elliptic}(a).
Like in the linear polarization case studied in Sec.~\ref{sec:numerical}, the FLZ theory well describes the numerical exact solution except for the vicinity of $a=\aACn{1}$, where the adiabatic-impulse approximation is not valid.
The oscillation of $\Pup$ in $a_0>\aACn{1}$ is due to the FLZ interferometry, and thus we have quantitatively elucidated the stripe-shaped pattern for $1<b<3$ in Fig.~\ref{fig:mask2}(b).
Note that, unlike the linear polarization case, we can implement the FLZ theory almost fully analytically in nearly-circular elliptic polarizations except for the numerical integration in obtaining $\Phi_{m,l}$ from analytically obtained quasienergies.

%#####

%#####
A similar analysis works for $0<b<1$ as well.
For this case, the Floquet-state hybridization occurs between $\alpha=(1,l+1)$ and $(2,l)$ as seen in Fig.~\ref{fig:quasi_elliptic}(b).
Namely, the first avoided crossing is caused by the 1-photon resonance.
Like in $1<b<3$ discussed above,
we can perform perturbation-theory analyses for these pairs of Floquet states, obtaining the avoided crossing point, quasienergies and their gap.
We plot the resulting $\Pup$ calculated by the FLZ theory in Fig.~\ref{fig:PLZS_elliptic} together with the exact numerical results.
Here again, we obtain a quantitatively good agreement between these results even though we resorted to perturbation-theory approximations.

\section{Justification of the FLZ theory}\label{app:justificaton}

\subsection{Relation to the semiclassical limit}\label{eq:semiclassical}\label{app:semiclassical}
Here we show that the limit of $\nu\to0$ formally corresponds to the semiclassical limit $\hbar\to0$.
For this purpose, we rewrite Eq.~\eqref{eq:IFSevol}, highlighting the approach to adiabaticity.
Let us define $\hat{a}(t)=a_0 \exp[-(t/T)^2]$, which coincides with $a(t)$ at $\nu=1$ and satisfies $a(t)=\hat{a}(t/\nu)=\hat{a}(\adi t)$.
Here, $\adi\equiv 1/\nu$ is the adiabaticity parameter~\cite{Davis1976}, and the $\adi\to0$ ($\nu\to\infty$) limit corresponds to the infinitely slowly varying envelope function.
Using $\hat{a}$ and introducing the rescaled time $s\equiv\adi t$, we rewrite Eq.~\eqref{eq:IFSevol} as
\begin{align}\label{eq:taueq}
	i \adi \frac{d \hat{c}_\alpha(s)}{ds} = \sum_\beta \mathcal{H}_{\alpha\beta}(\hat{a}(s))\hat{c}_\beta(s)
\end{align}
with $\hat{c}_\alpha(s)\equiv c_\alpha(t/\adi)$.
Equation~\eqref{eq:taueq} shows that the adiabatic limit $\adi\to0$ is formally equivalent to the semiclassical limit $\hbar\to0$.
Considering that the Landau-Zener-type transition derives from the series expansion for $\hbar$ and its resummation~\cite{Drese1999,Berry1990}, we naturally expect that our FLZ theory works well when $\adi\ll1$, i.e., $\nu\gg1$.

The above argument is, at least, consistent with our numerics in Fig.~\ref{fig:widthdep}, where the FLZ theory works well for $\nu\ge2$ but not for $\nu=1$.
However, characterizing the precise threshold, which should be model-dependent, and showing why it is about 2 in the present model remain open for future work.

\subsection{Transitions to distant IFSs}\label{app:transitions}
The transfer matrix $\mT_n$ only mixes pairs of IFSs that have the nearest quasienergies
although the mixings can, in principle, happen between other IFS pairs that are more distant in the quasienergy.
These transitions neglected in our FLZ theory could be nonnegligible when the pulse width $\nu$ becomes very small.
However, we show here that these transitions are not relevant in the present model.

To study this possibility quantitatively, we rewrite Eq.~\eqref{eq:IFSevol} so that the $\nu$-dependence is evident:
\begin{align}
	i\frac{d\tilde{c}_\alpha(\tau)}{d\tau} &= \sum_\beta \tilde{\mathcal{H}}_{\alpha\beta}(\tilde{a}(\tau)) \tilde{c}_\beta(\tau),\label{eq:IFSevol2}\\
	\tilde{\mathcal{H}}_{\alpha\beta}(\tilde{a}(\tau)) &\equiv \nu T\delta_{\alpha\beta}\epsilon_\alpha(\tilde{a}(\tau))-i\frac{d\tilde{a}}{d\tau}\mathcal{G}_{\alpha\beta}(\tilde{a}(\tau)).\label{eq:FHam2}
\end{align}
Here, $\tau\equiv \frac{t}{\nu T}$ is a dimensionless time, $\tilde{c}_\alpha(\tau)=c_\alpha(\frac{t}{\nu T})$, and $\tilde{a}(\tau)=a(\frac{t}{\nu T})=a_0\exp(-\tau^2)$.
In this representation, $\nu$ effectively rescales the quasienergy as the first term on the right-hand side of Eq.~\eqref{eq:FHam2}.
For an IFS pair $(\alpha,\beta)$, their transition is negligible if their effective quasienergy difference $\nu T| \epsilon_\alpha -\epsilon_\beta|$ is much larger than their coupling $|\frac{d\tilde{a}}{d\tau}\mathcal{G}_{\alpha\beta}(\tilde{a}(\tau))|$.
Our IFS theory neglects the pair transitions for $|\epsilon_\alpha-\epsilon_\beta|>\omega$, and this treatment is justified if
\begin{align}\label{eq:criterion}
	\nu \gg \nu_c \equiv \frac{1}{T}\max_\tau \max_{\substack{\alpha,\beta \\ |\epsilon_\alpha-\epsilon_\beta|>\omega}}
	%\left|\frac{d\tilde{a}}{d\tau}\mathcal{G}_{\alpha\beta}(\tilde{a}(\tau))\right|.
	\left|\frac{d\tilde{a}}{d\tau}\frac{\mathcal{G}_{\alpha\beta}(\tilde{a}(\tau))}{\epsilon_\alpha(a(\tau))-\epsilon_\beta(a(\tau))}\right|.
\end{align}

\begin{figure}
	\includegraphics[width=\columnwidth]{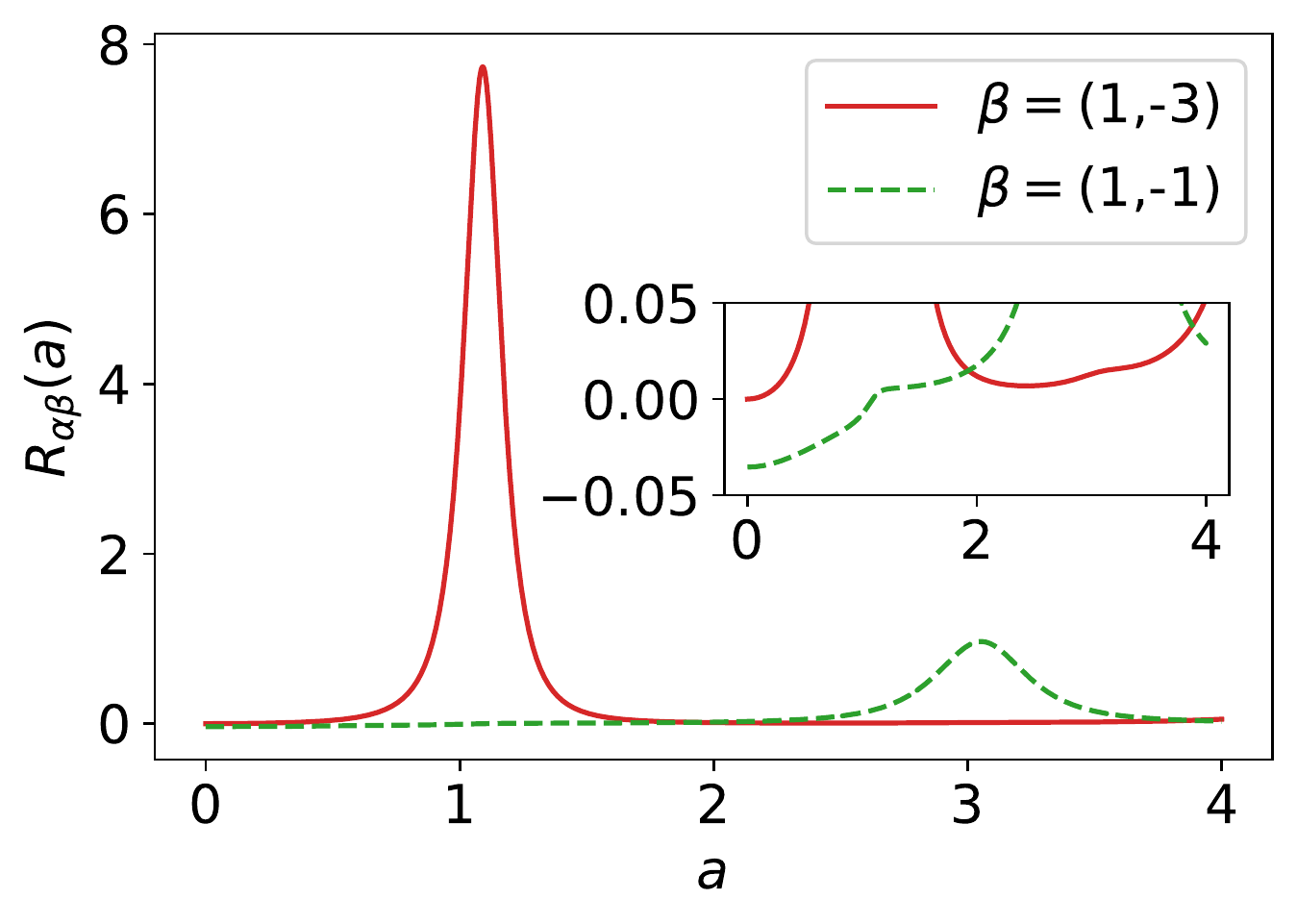}
	\caption{The ratio $\mathcal{G}_{\alpha\beta}(a)/\{T[\epsilon_\alpha(a)-\epsilon_\beta(a)]\}$ for $\alpha=(2,0)$ and $\beta=(1,-3)$ (solid) and $(1,-1)$ (dashed).
	Here we set $b=2.5$ as in Figs.~\ref{fig:quasi} and \ref{fig:widthdep}.
	}
	\label{fig:gmat}
\end{figure}

%##
Now we numerically confirm that $\nu_c$ is so small that the transitions to distant IFS are negligible.
Since $|d\tilde{a}/d\tau|<0.74$, we focus on the ratio
\begin{align}
	R_{\alpha\beta}(a) \equiv \frac{\mG_{\alpha\beta}(a)}{[\epsilon_\alpha(a)-\epsilon_\beta(a)]T}
\end{align}
and verify $|R_{\alpha\beta}(a)|\ll1$ for $|\epsilon_\alpha(a)-\epsilon_\beta(a)|>\omega$.
This ratio is shown in Fig.~\ref{fig:gmat} for $\alpha=(2,0)$ and $b=2.5$, which we mainly argued in the main text.
The ratio $R_{\alpha\beta}(a)$ shows large peaks at $a=\aACn{1}=1.09$ for $\beta=(1,-3)$ and at $a=\aACn{2}=3.05$ for $\beta=(1,-1)$, which correspond to $|\epsilon_\alpha(a)-\epsilon_\beta(a)|<\omega$ and are responsible for the transitions incorporated by the transfer matrices $\mT_1$ and $\mT_2$, respectively (see Fig.~\ref{fig:quasi}).
Except for these peaks, $|R_{\alpha\beta}(a)|\lesssim0.05$ for $0\le a\le 4$ as shown in the inset ($|R_{\alpha\beta}|$ is much smaller for other $\beta$'s not shown in the figure).
This means $\nu_c<0.05$ for $a_0\le4$ considered in the main text.

Recall that the FLZ theory starts to fail when $\nu\sim1$ as shown in Fig.~\ref{fig:widthdep}.
However, as $\nu_c<0.05$, Eq.~\eqref{eq:criterion} still remains true.
Thus, we conclude that the transition to distant IFSs is not the main reason for the FLZ's failure at the ultrashort pulse as short as $\nu\sim1$.

\end{document}